\documentclass[11pt]{article}

\usepackage[preprint]{acl}
\usepackage{amsmath}
\usepackage{times}
\usepackage{latexsym}
\usepackage[T1]{fontenc}
\usepackage{microtype}
\usepackage{hyperref}
\usepackage{url}
\usepackage{booktabs}
\usepackage{graphicx}
\usepackage{wrapfig}
\usepackage{amsmath}
\usepackage{multirow}
\usepackage[table]{xcolor}
\usepackage{float}

\DeclareMathOperator*{\argmin}{arg\,min}
\usepackage{amsmath}
\usepackage{amsfonts}
\usepackage{mathrsfs}
\usepackage{caption}
\usepackage{makecell}
\usepackage{graphicx}
\usepackage{multirow}
\usepackage[skins]{tcolorbox}
\usepackage{enumitem}

\usepackage[utf8]{inputenc}

\usepackage{microtype}

\usepackage{inconsolata}

\usepackage{graphicx}

\usepackage{lineno}

\definecolor{darkblue}{rgb}{0, 0, 0.5}
\hypersetup{colorlinks=true, citecolor=darkblue, linkcolor=darkblue, urlcolor=darkblue}
\newcommand{\cellgq}{\cellcolor{gray!15}}   
\newcommand{\cellgl}{\cellcolor{gray!30}}   

\title{RL-Index: Reinforcement Learning for Retrieval Index Reasoning}

\author{
\normalfont
Yongjia Lei$^\heartsuit$ \quad 
Nedim Lipka$^\clubsuit$ \quad 
Zhisheng Qi$^\heartsuit$ \quad 
Utkarsh Sahu$^\heartsuit$ \quad
Yuchen Zhuang$^\diamondsuit$ \\
Wenqi Shi$^\spadesuit$ \quad 
Koustava Goswami$^\clubsuit$ \quad
Franck Dernoncourt$^\clubsuit$ \quad 
Ryan A. Rossi$^\clubsuit$ \quad 
Yu Wang$^\heartsuit$ \\
\\
$^\heartsuit$University of Oregon \quad $^\clubsuit$Adobe Research \quad $^\diamondsuit$Google DeepMind \\ $^\spadesuit$University of Texas Southwestern Medical Center\\
\texttt{\{yongjia, zhisheng.qi, utkarsh, yuwang\}@uoregon.edu}\quad\texttt{Wenqi.Shi@UTSouthwestern.edu}\\
\texttt{\{lipka, koustavag, dernonco, ryrossi\}@adobe.com} \quad \texttt{night.yuchen@gmail.com}
}

\begin{document}
\maketitle
\begin{abstract}
Retrieving external knowledge is crucial for real-world tasks but remains difficult when queries and relevant knowledge are linked by implicit reasoning (e.g., shared theorems or coding logic). Existing methods rely mainly on query-side reasoning, leading to high online latency and underutilizing the reasoning semantics within the knowledge corpus.
In this paper, we propose \textbf{RL-Index}, an indexing framework that formulates \emph{retrieval index reasoning} as a reinforcement learning problem. Instead of performing reasoning at query time, RL-Index shifts reasoning to the indexing stage by augmenting documents with LLM-generated rationales that explicitly encode the latent query-knowledge relationship. To optimize the quality of these rationales, we employ Group Relative Policy Optimization (GRPO) and use retrieval similarity as a proxy reward signal, enabling direct optimization of indexing decisions for retrieval effectiveness.
Extensive experiments on the BRIGHT benchmark demonstrate that RL-Index consistently improves both retrieval and downstream question-answering performance, while significantly reducing online inference latency. Moreover, the learned rationale augmentation generalizes across diverse retrievers and generators, highlighting its robustness as a plug-and-play indexing strategy across different retrieval systems\footnote{Our code \href{https://github.com/Yoega/RL-Index}{https://github.com/Yoega/RL-Index}.}
\end{abstract}

\begin{figure}[t!]
    \centering
    \includegraphics[width=1\linewidth]{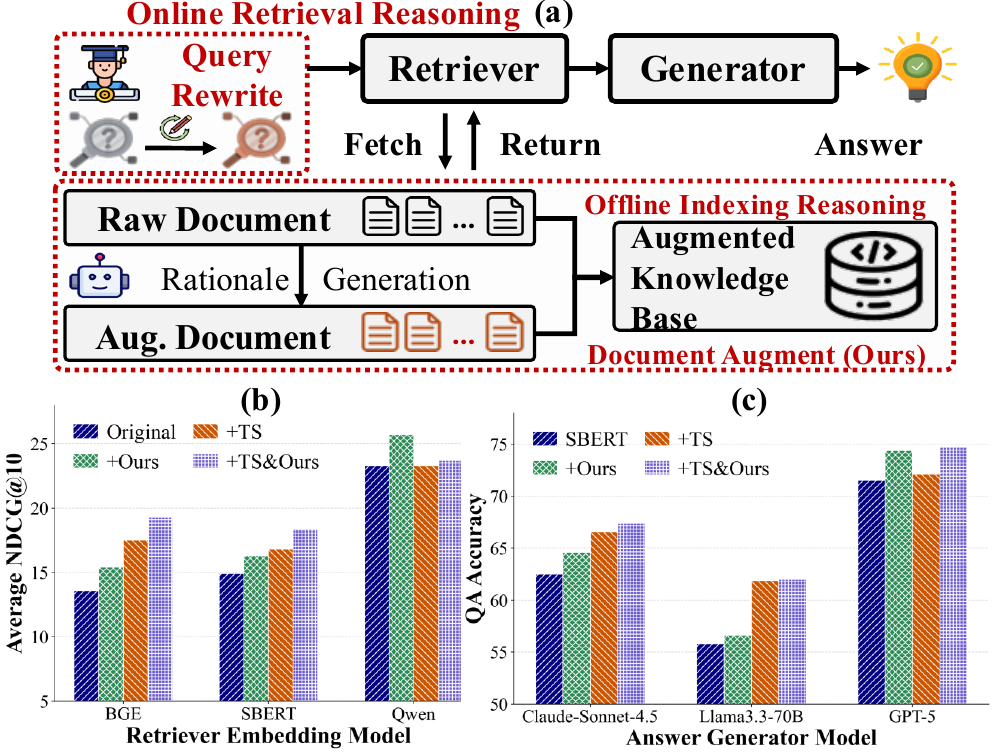}
    \caption{\textbf{(a)} Online retrieval reasoning by query rewriting (e.g., TongSearch (TS)~\citep{qin2025tongsearch}) and Offline indexing reasoning by document augmentation (RL-Index (Ours)).
    \textbf{(b)-(c)} Both TS and RL-Index improve \textbf{(b)} retrieval and \textbf{(c)} QA performance, and their gains are further compounded when combined together.}
    \label{fig-motivation}
    \vspace{-3.5ex}
\end{figure}

\vspace{-1ex}
\section{Introduction}
\vspace{-1ex}
Retrieving relevant knowledge to augment downstream tasks (e.g., question answering, fact checking, and text generation)~\cite{guu2020retrieval,wang2024knowledge,han2024retrieval,qi2026rigorizing} has become fundamental to applications ranging from scientific discovery and cybersecurity to social wellness~\cite{wu2024stark,lei2025mixture,rahman2024retrieval,zhangpersonalization} and knowledge/memory management in LLM-powered agents~\cite{wu2025memory,huang2026rethinking}. However, traditional retrieval methods based on semantic embeddings or lexical matching (e.g., TF-IDF and BM25) primarily capture surface-level similarity, making them ineffective when knowledge relevance depends on underlying logical relationships rather than textual overlap~\cite{shao2025reasonir,zhuang2025rank,hongjinbright}. 
For example, a mathematical query and relevant solutions may rely on the same underlying theorem despite differing surface expressions.~\cite{hongjinbright,alexander2026semantic}. 
Similarly, scientific and legal documents often describe underlying events without explicitly naming the queried concept or mechanism (e.g., ``breach of fiduciary duty''~\cite{paul-etal-2025-il}).
%

To capture these complex logical relations, existing retrieval methods generate intermediate rationales, which can be broadly categorized into online query and offline index reasoning (Figure~\ref{fig-motivation}\textbf{(a)}). Online query reasoning infers latent user intent through LLM-generated rationales
~\cite{jinsearch,jagerman2023query,zhang2025query}. However, it incurs substantial query-time latency and limits reasoning to the query itself, leaving the rich reasoning semantics in the knowledge corpus underutilized~\cite{chen2025enrichindex,lee2025imagine}. Offline index reasoning addresses these issues by enriching documents with query-oriented rationales in advance~\cite{gospodinov2023doc2query}. For example, SPIKE~\cite{lee2025imagine} synthesizes potential user intents and EnrichIndex~\cite{chen2025enrichindex} augments documents with multiple views, such as summaries, purposes, and QA pairs. However, both approaches rely on expensive closed-source LLMs and prompt-engineering that is neither optimized for retrieval objectives nor tailored to domain-specific knowledge. Doc2Query~\cite{nogueira2019document} predicts potential queries for a document using supervised learning, without learning the underlying rationales that relate queries to documents.

To address the above limitations, we propose RL-Index, an agentic indexing framework that formulates retrieval index reasoning as an offline optimization problem. Rather than reasoning at query time, RL-Index employs an LLM-powered agent to enrich documents with logical rationales.
To further improve rationale quality, we optimize the indexing agent with Group Relative Policy Optimization (GRPO)~\cite{shao2024deepseekmath,guo2025deepseek}, using incremental document relevance as the reward to directly align generated rationales with retrieval objectives and domain-specific knowledge. In Figure~\ref{fig-motivation} and Table~\ref{tab-online}, RL-Index improves retrieval and downstream QA performance without additional online latency, while remaining complementary to query-side reasoning. To summarize, our contributions are:
\vspace{-1ex}
\begin{itemize}[leftmargin=*, itemsep=0pt]
    \item We introduce an indexing framework that tackles reasoning-intensive retrieval from an offline perspective, improving retrieval of implicitly relevant documents while simultaneously reducing inference latency.
    
    \item In particular, we are the first to formulate document augmentation as a reinforcement learning problem. We optimize an open-source LLM with GRPO to reason over documents with a simple yet effective reward design that aligns the training objective with the desired reasoning behavior.
    
    \item Extensive experiments on BRIGHT across multiple retrievers and LLMs demonstrate consistent retrieval and QA performance improvements with strong transferability, substantial efficiency gains, and interpretable rationale analysis.
\end{itemize}

\section{Related Work}\label{sec-relatedwork}
\vspace{-1.3ex}
\textbf{Reasoning-Intensive Knowledge Retrieval.} Reasoning-intensive retrieval~\citep{hongjinbright,yao2023react} focuses on queries that require deeper reasoning to uncover complex logical relationships (e.g., multi-hop connections~\citep{xiong2020answering,trivedi-etal-2023-interleaving, han2025reasoning} or shared mathematical principles~\citep{alexander2026semantic}) instead of simple lexical and semantic matching.
Thus, first-stage retrieval is often the bottleneck, as relevant documents are usually not surface-matched to the query, or the key evidence is distributed across the corpus~\citep{trivedi-etal-2023-interleaving}. Recent work has improved retrievers with stronger dense representations~\citep{shao2025reasonir, das2025rader} and hybrid sparse-dense scoring~\citep{kalra2025mor}. 
However, these approaches still rely on the original query and document form, which may not explicitly expose the latent rationale needed for retrieval.

\noindent\textbf{Online Retrieval Reasoning by Query Rewriting.} This line of work addresses the lack of rationale by injecting explicit reasoning into queries at inference time. Existing methods interleave reasoning and retrieval across multiple turns with interactive feedback~\citep{jinsearch,trivedi2023interleaving,yao2023react}, or infuse queries with explicit rationales~\citep{lei-etal-2025-thinkqe,qin2025tongsearch}. While they achieve strong performance on reasoning-intensive benchmarks~\citep{hongjinbright,wen2020answer}, they incur substantial online latency and remain limited when evidence resides in documents rather than queries~\citep{chen2025enrichindex}.

\noindent\textbf{Offline Reasoning by Document Augmentation.}
To reduce inference-time overhead, another category of work shifts reasoning from online to offline indexing by augmenting documents. 
Some methods augment documents from single perspectives, e.g., pseudo queries and summaries~\cite{gospodinov2023doc2query,jeong2021unsupervised}. Others infuse richer perspectives, e.g., synthetic user scenarios~\citep{lee2025imagine} and combinations of enrichment signals (i.e., summary, purpose, and QA pairs)~\citep{chen2025enrichindex}. Compared with online query reasoning, offline augmentation provides a latency-quality trade-off as retrieval can remain a single pass at inference. Our work follows this line and formulates document augmentation as a policy-learning problem where a document augmenter learns how to generate augmented documents that make latent rationale explicit, improving retrieval.

\begin{figure*}[t!]
    \centering
\includegraphics[width=1\linewidth]{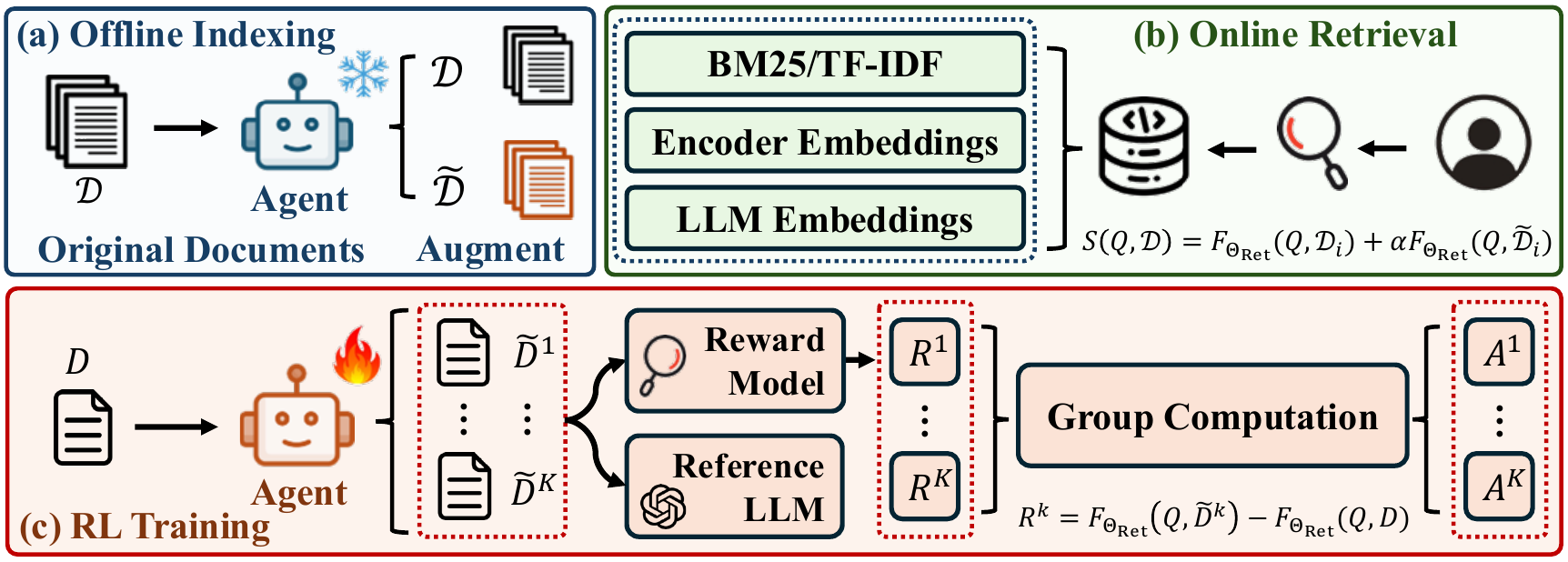}
    \vspace{-3.5ex}
    \caption{Overview of RL-Index framework.
    \textbf{(a)} In offline indexing, an RL-trained agent augments documents $\mathcal{D}$ into $\widetilde{\mathcal{D}}$.
    \textbf{(b)} In online retrieval, augmented corpus $\widetilde{\mathcal{D}}$ is used for improved document retrieval.
    \textbf{(c)} RL indexer is trained via GRPO, with rewards as incremental relevance gain of augmented over original documents to the query.} 
    \label{fig-framework}
    \vspace{-3ex}
\end{figure*}


\section{Framework}\label{sec-framework}
\vspace{-1ex}
\subsection{Math Notations and Overview}
Let $\mathcal{D}=\{\mathcal{D}_i\}_{i=1}^{|\mathcal{D}|}$ denote a document corpus, $Q$ a user query, and $\mathcal{D}^{*}\subseteq\mathcal{D}$ the relevant documents. A conventional retriever $F_{\boldsymbol{\Theta}_{\text{Retriever}}}$ scores each document by $F_{\boldsymbol{\Theta}_{\text{Retriever}}}(Q,\mathcal{D}_i)$ and returns the top-$K$ results. However, implicit evidence is often not sufficiently expressed in the original documents. To address this limitation, we formulate retrieval enhancement as an \emph{offline indexing-time reasoning problem}, shifting reasoning from online query processing to document augmentation. In Figure~\ref{fig-framework}(a), an LLM-based indexing agent generates a rationale for each document,
$\widetilde{\mathcal{D}}_i = F_{\boldsymbol{\Theta}_{\text{Indexer}}}(\mathcal{D}_i;P)$,
where prompt $P$ guides the agent to expose latent logical relations relevant to future queries. Applying the indexer to the entire corpus yields
$\widetilde{\mathcal{D}}=\{\widetilde{\mathcal{D}}_i\}_{i=1}^{|\mathcal{D}|}$,
with each pair $(\mathcal{D}_i,\widetilde{\mathcal{D}}_i)$ forming an enriched index.


During online retrieval in Figure~\ref{fig-framework}(b), no additional LLM reasoning is required. Given $Q$, the retriever jointly scores each document and its rationale and selects the top-$K$ candidates:
\begin{equation}
\widehat{\mathcal{D}}(Q)
=
\operatorname{TopK}_{\mathcal{D}_i\in\mathcal{D}}
F_{\boldsymbol{\Theta}_{\text{Retriever}}}
(
Q,\mathcal{D}_i,\widetilde{\mathcal{D}}_i
).
\end{equation}
The retrieved evidence is then passed to the answer generator:
$\widehat{A}=F_{\boldsymbol{\Theta}_{\text{Answer}}}
(Q,\widehat{\mathcal{D}}(Q))$.
The retriever can be instantiated with lexical methods (e.g., BM25 or TF-IDF), embedding-based models (e.g., SBERT or BGE), or LLM-derived embedding models (e.g., Qwen). 
By jointly considering the query $Q$, the original document $D_i$, and the generated rationale $\widetilde{D}_i$, the retriever can better capture complex logical relations between user intent and the implicit evidence contained in either the original or the augmented documents.


In the following sections, we introduce the Agentic Indexer in Section~\ref{sec-rationale}, present its GRPO-based optimization in Section~\ref{sec-rl}, and describe the integration of original and augmented documents into online retrieval in Section~\ref{sec-index}.

\subsection{Offline Rationale Generation}\label{sec-rationale}
\vspace{-1ex}
Raw documents often lack explicit links to user intents, hindering retrieval for queries requiring complex and implicit reasoning. To address this, we introduce an indexing approach with offline rationale generation to externalize the logical connections between user intent and document knowledge. The generated rationales consist of two components:

\textbf{Thematic Synthesis (Key Points).} Rather than producing a surface-level summary, our model distills each document into a compact set of core propositions~\citep{chen2024dense}. These propositions characterize the document from diverse perspectives and extract global facts that can satisfy potential user needs. 

\textbf{Functional Alignment (Explanations).} Building on the key points, the model then articulates how those propositions satisfy potential user needs. This stage links document
content to the retrieval intent desired by the potential incoming queries. By constraining
explanations to be derived solely from extracted key points, we ensure rationale traceability
to verified document content.

We implement offline rationale generation with a structured prompt that produces both key points and corresponding explanations (see next). The resulting rationale pairs are used for computing query-document similarity used for reward computing
during GRPO-based training in Section~\ref{sec-rl} and for retrieval at inference in Section~\ref{sec-index}.

\begin{tcolorbox}[
    enhanced,
    colback=white,
    colframe=green!40!black,
    arc=10pt,
    fonttitle=\bfseries\scriptsize,
    boxrule=1.5pt,
    title=\textbf{Prompt for Generating Rationale to Augment Documents},
    colbacktitle=green!40!black,
    attach boxed title to top left={xshift=5mm, yshift=-3mm},
    boxed title style={colframe=green!40!black, sharp corners=southwest}
]
    \small
    You are an advanced language model specializing in knowledge extraction and user need modeling. Your task is to extract hypothetical user scenarios from a given document, ensuring that the generated information needs reflect the document's overall insights and knowledge, rather than isolated details.

    \textbf{Content:}
    \begin{itemize}[noitemsep, topsep=2pt, leftmargin=15pt]
        \item \textbf{Key Points:} Summarize the core concepts, insights, or knowledge presented.
        \item \textbf{Explanations:} Explain how the document fulfills the hypothetical user need, ensuring that explanations are generalized and conceptual rather than overly detailed.
    \end{itemize}

    \vspace{0.5em}
    \textbf{Input:} $<$Document$>$ \\
    \{doc\} \\
    \textbf{Output:}
    \label{tab-prompt}
\end{tcolorbox}

\subsection{Reinforcement Learning Optimization}\label{sec-rl}
Inspired by DeepSeek-R1~\citep{Guo_2025}, we employ Group Relative Policy Optimization (GRPO) to optimize our LLM-powered Agentic Indexer for document rationale generation. 
In our framework, the LLM-powered Agentic Indexer takes a document $D$ as input and generates a set of rationale-augmented versions. For each document, we sample a group of $K$ augmented candidates $\{\widetilde{D}^k\}_{k=1}^{K}$ from an old policy $\pi_{\boldsymbol{\Theta}_{\text{old}}}$. The current policy $\pi_{\boldsymbol{\Theta}}$ is optimized to assign higher probability to augmented documents with larger relative advantages within each group, while remaining close to the reference policy:
\vspace{-1ex}
\begin{equation}
\small
\begin{aligned}
&\boldsymbol{\Theta}^* = \argmin_{\boldsymbol{\Theta}} \mathbb{E}_{(Q, D)\sim \mathbb{Q}\times \mathbb{D}, \{\widetilde{D}^k\}_{k=1}^K \sim \pi_{\boldsymbol{\Theta}_{\text{old}}}(\cdot|D)} \frac{1}{K} \sum_{k=1}^{K}\\
&[\min (\rho A^k,\ \text{clip}( \rho, 1\pm\epsilon) A^k) - \beta\, \mathrm{KL}(\pi_{\boldsymbol{\Theta}} \,\|\, \pi_{\mathrm{ref}})]
\end{aligned}
\label{eq-grpo}
\end{equation}
where the ratio $\rho=\frac{\pi_{\boldsymbol{\Theta}}(\widetilde{D}^k|D)}{\pi_{\boldsymbol{\Theta_{\text{old}}}}(\widetilde{D}^k|D)}$ compares the updated policy to the reference policy for the same augmented document, and ${A}^k$ is the relative advantage computed by normalizing the rewards within the augmented document group $\{R^k\}_{k=1}^{K}$: $A^k = \frac{R^k - \text{MEAN}(R^1, \dots, R^k)}{\text{STD}(R^1, \dots, R^K) + \delta}$ where $\delta$ is a constant used to avoid the term divided by zero and $\text{MEAN}/\text{STD}$ denote the average and standard deviation term. The clip function in Eq.~\eqref{eq-grpo} is used to constrain the importance ratio $\rho$ within the range $[1-\epsilon, 1+\epsilon]$. This mechanism prevents the policy from changing too drastically in a single update, which ensures training stability and prevents the model from collapsing during the reinforcement learning~\citep{JoHn2017prox}. 
The KL term also acts as a regularizer that keeps the model grounded near the reference policy.

Most prior work~\citep{jiang2025deepretrieval,zhuang2022reinforcement} defines rewards using retrieval metrics such as nDCG or Recall. However, in our setting, the action corresponds to \emph{document-side augmentation}. Directly optimizing such metrics would require re-augmenting the corpus and re-running retrieval after every policy update, which is computationally prohibitive. Following TongSearch~\citep{qin2025tongsearch}, we instead adopt a lightweight \emph{similarity-gain reward}. For each training pair $(Q, D)$, the reward for an augmented document $\widetilde{D}^k$ is defined as $R^k = F_{\boldsymbol{\Theta}_{\text{Retriever}}}(Q, \widetilde{D}^k) - F_{\boldsymbol{\Theta}_{\text{Retriever}}}(Q, D)$ where $F_{\boldsymbol{\Theta}_{\text{Retriever}}}(\cdot,\cdot)$ denotes the retrieval score between any query-document pair. This reward is computationally efficient, requiring only two embedding forward passes per sample, while still providing a strong retrieval-oriented signal. 
In this work, we implement $F_{\boldsymbol{\Theta}_{\text{Retriever}}}(\cdot,\cdot)$ as cosine similarity between query and document embeddings computed using a fixed embedding model.

\subsection{Online Rationale-augmented Retrieval}\label{sec-index}
After training the RL-based Rationale-Augmented Indexer, we construct two parallel dense indices over the corpus. For each document $D$, we maintain two representations: the original document $D$ and its rationale-augmented version $\widetilde{D}$, both encoded using the same fixed embedding model. At query time, a query $Q$ is embedded using the same encoder and independently matched against both indices. The final retrieval score is $S(Q, D) = F_{\boldsymbol{\Theta}_{\text{Retriever}}}(Q, D) + \alpha \, F_{\boldsymbol{\Theta}_{\text{Retriever}}}(Q, \widetilde{D})$ where $\alpha$ controls the augmented view contribution and is set to 1 by default. 
Documents are ranked by retrieval score $S(\cdot)$, preserving original evidence while leveraging augmented rationales to bridge latent gaps. 


\begin{table*}[t!]
\centering
\small
\setlength{\tabcolsep}{3.7pt}
\begin{tabular}{l|ccccc|cccc|ccc|cc}
\toprule
\multirow{2.5}{*}{\textbf{Model}} & \multicolumn{5}{c|}{\textbf{Natural Language}} 
 & \multicolumn{4}{c|}{\textbf{Code}} 
 & \multicolumn{3}{c|}{\textbf{Math}} 
 & \multirow{2.5}{*}{\textbf{Avg.}} & \multirow{2.5}{*}{\textbf{Improv.}} \\
\cmidrule(r){2-13}
& \textbf{Bio.} & \textbf{Earth.} & \textbf{Econ.} & \textbf{Psy.} & \textbf{Sus.}
& \textbf{Rob.} & \textbf{Stack.} & \textbf{Leet.} & \textbf{Pony}
& \textbf{Aops} & \textbf{TheoQ.} & \textbf{TheoT.}
&  &  \\
\midrule
BGE 
& 11.7 & 24.4 & 16.4 & 17.4 & 13.1 & 11.7 & 10.6 & 26.7 & 5.7 & 6.0 & 13.0 & 6.9 & 13.6 & -- \\
+SPIKE$^*$ 
&13.0 &24.4 &13.3 &18 &13.5 &12.2 &13.1 &26.0 &7.7 &5.5 &12.7 &8.0 &14.0 &+3.0\%\\
+SPIKE & 13.2 & 26.4 & \textbf{17.0} & 18.1 & 13.2
& 11.5 & 13.3 & \textbf{27.1} & 6.4
& 4.8 & 13.0 & 8.5
& 14.4 & +5.9\% \\
Doc2Query & 9.5  & 23.9 & 15.5 & 17.3 & 13.1 & 10.6 & 10.5 & 25.6 & 3.8  & \textbf{7.0} & 12.6 & 4.5 & 12.8 & -5.9\% \\
\cellcolor{gray!30}+RL-Index 
& \cellcolor{gray!30}\textbf{14.1} & \cellcolor{gray!30}\textbf{27.2} & \cellcolor{gray!30}16.9 & \cellcolor{gray!30}\textbf{18.9} & \cellcolor{gray!30}\textbf{14.0} & \cellcolor{gray!30}\textbf{14.0} & \cellcolor{gray!30}\textbf{14.0} & \cellcolor{gray!30}26.0 & \cellcolor{gray!30}\textbf{10.5} & \cellcolor{gray!30}5.9 & \cellcolor{gray!30}\textbf{13.9} & \cellcolor{gray!30}\textbf{9.6} & \cellcolor{gray!30}\textbf{15.4} & \cellcolor{gray!30}\textbf{+13.2\%} \\
\hline
SBERT 
& 15.2 & 20.4 & 16.6 & \textbf{22.7} & 15.3 & 8.2 & 11.0 & 26.4 & 7.0 & 5.3 & 20.0 & 10.8 & 14.9 & -- \\
+SPIKE$^*$ 
& 16.9 & 22.0 & 13.3 & 20.0 & 15.3 & 9.6 & 13.2 & 26.4 & 8.1 & 4.6 & 19.2 & 11.3 & 15.0 & +0.7\% \\
+SPIKE
& \textbf{18.2} & \textbf{23.1} & 17.9 & 21.3 & 15.5
& 9.0 & 13.4 & 26.7 & 8.1
& 5.4 & 19.3 & 11.2
& 15.8 & +6.0\% \\
Doc2Query &17.1 &20.3 &17.8 &22.7 &\textbf{19.2} &\textbf{11.0} &14.3 &26.8 &7.8 &\textbf{6.0} &18.7 &8.4 &15.8 &+6.0\%\\
\cellcolor{gray!30}+RL-Index 
& \cellcolor{gray!30}15.7 & \cellcolor{gray!30}22.5 & \cellcolor{gray!30}\textbf{18.9} & \cellcolor{gray!30}21.5 & \cellcolor{gray!30}16.1 & \cellcolor{gray!30}10.5  & \cellcolor{gray!30}\textbf{14.7} & \cellcolor{gray!30}\textbf{28.3} & \cellcolor{gray!30}\textbf{8.5} & \cellcolor{gray!30}5.4 & \cellcolor{gray!30}\textbf{20.9} & \cellcolor{gray!30}\textbf{13.1} & \cellcolor{gray!30}\textbf{16.3} & \cellcolor{gray!30}\textbf{+9.4\%} \\
\hline
Qwen 
& 29.9 & 39.6 & 17.7 & 24.4 & 20.3 & 13.2 & 21.2 & 25.5 & 12.4 & 14.4 & 27.8 & 32.9 & 23.3 & -- \\
+SPIKE$^*$ 
& \textbf{32.8} & 36.6 & 18.3 & 25.7 & 24.9 & 14.8 & 21.6 & 25.7 & 16.7 & 12.9 & 26.6 & 28.8 & 23.8 & +2.2\% \\
+SPIKE 
& 32.4 & \textbf{41.2} & \textbf{23.7} & 25.7 & 24.7 & 16.0 & \textbf{23.7} & 26.3 & 16.7 & 12.5 & 27.1 & 31.0 & 25.1 & +7.7\% \\
Doc2Query &18.6 &32.9 &18.7 &20.8 &16.4 &13.6 &15.7 &23.6 &15.2 &14.3 &23.4 &12.2 &18.8 &-19.3\% \\
\cellcolor{gray!30}+RL-Index 
& \cellcolor{gray!30}29.8 & \cellcolor{gray!30}39.7 & \cellcolor{gray!30}21.9 & \cellcolor{gray!30}\textbf{27.8} & \cellcolor{gray!30}\textbf{26.7} & \cellcolor{gray!30}\textbf{16.6} & \cellcolor{gray!30}22.1 & \cellcolor{gray!30}\textbf{28.3} & \cellcolor{gray!30}\textbf{17.0} & \cellcolor{gray!30}\textbf{16.0} & \cellcolor{gray!30}\textbf{28.5} & \cellcolor{gray!30}\textbf{33.6} & \cellcolor{gray!30}\textbf{25.7} & \cellcolor{gray!30}\textbf{+10.3\%} \\

\bottomrule
\end{tabular}
\vspace{-1ex}
\caption{Comparison of retrieval performance (nDCG@10) using document augmentation with rationales from existing offline indexing baselines and our proposed RL-Index. SPIKE$^*$/SPIKE denote our reproduced/originally reported versions of the baseline. RL-Index consistently achieves the best performance.}
\label{tab-unified-llama}
\vspace{-2ex}
\end{table*}

\section{Experiment}\label{sec-experiment}
\vspace{-1ex}
\subsection{Experimental Setup}
\textbf{Training Datasets.} We employ training data \textbf{V2}~\citep{qin2025tongsearch}, which includes around 30K query-document pairs with logical relations, containing biology, chemistry, code review, CS, earth science, economics, math, physics, robotics. Details are in Appendix~\ref{app-dataset}.

\textbf{Evaluation Datasets and Metrics.} We use BRIGHT~\citep{hongjinbright}, a benchmark for reasoning-intensive retrieval containing 1,384 real-world queries spanning 12 datasets across diverse domains. Following~\citet{lee2025imagine}, we report nDCG@10 for all evaluations.

\noindent\textbf{Baseline Comparisons.} 
To evaluate the effectiveness of offline rationale generation, we compare RL-Index with \textsc{SPIKE}~\citep{lee2025imagine} and Doc2query~\citep{nogueira2019document}. SPIKE synthesizes scenario-aware document expansions by supervised fine-tuning a small LLM using examples derived from GPT-4o. Doc2query predicts potential user queries by supervised learning on query-document pairs.
In contrast, RL-Index optimizes document augmentation via retrieval-based rewards. We compare both of these offline reasoning methods across diverse retrievers (SBERT, BGE, Qwen~\citep{li2023towards, reimers2019sentence, xiao2024c}) and LLM-powered rationale generators (Llama3.2-3B-Instruct in Table~\ref{tab-unified-llama} and Qwen2.5-1.5B-Instruct~\citep{grattafiori2024llama, hui2024qwen2} in Table~\ref{tab-transfer-llm}).
We also compare against the state-of-the-art online query reasoning method, TongSearch-QR (TongSearch hereafter)~\citep{qin2025tongsearch}.

\noindent\textbf{Implementation Details.} Our LLM Agentic Indexer is trained on a single node with 4 NVIDIA H100-80G GPUs. We run GRPO with a per-device batch size of 16 or 8 with 16 rollouts per prompt $K=16$. Training lasts for 1000 optimization steps with a learning rate of $1\mathrm{e}{-6}$ and a KL coefficient $\beta$ of 0.008. All results are averaged over the final three checkpoints saved every 100 steps. RL rewards are computed using similarity gains from a training-time retriever. Evaluation is conducted with an inference-time retriever to assess deployment performance and transferability (Table~\ref{tab-transfer-retriever}-\ref{tab-transfer-llm}). Notably, the retriever used during training may differ from that used at inference to demonstrate the transferability of RL-Index. We fix $\alpha = 1$ for all experiments (see Appendix~\ref{app-alpha} for an ablation on this parameter).
%
%

\subsection{Overall Retrieval Effectiveness}
\vspace{-1ex}
Table~\ref{tab-unified-llama} reports results using Llama-3.2-3B-Instruct as the rationale generator under a unified training and evaluation setup. Across all three retrievers (BGE, SBERT, and Qwen), our RL-Index consistently achieves the best nDCG@10. For encoder-only models (SBERT and BGE), the improvements are stable across most sub-tasks and translate into clear average gains, indicating that rationale augmentation effectively enhances smaller retrievers.
Even for the large-scale decoder-only retriever (Qwen), our method also yields consistent improvements, suggesting that document-side augmentation can further complement stronger retrievers.
We use the pretrained SPIKE model from Hugging Face for inference. Due to unspecified generation parameters (e.g., max tokens, temperature), our reproduced results $\text{SPIKE}^{*}$ are slightly lower than reported. 
We also evaluate the Doc2Query model for document augmentation by generating synthetic queries. However, Doc2Query does not consistently improve reasoning-intensive retrieval. We hypothesize that directly appending synthetic queries to the original document introduces noise, reducing the quality of the document representation.
Overall, results demonstrate that our RL-trained Agentic Indexer generates more effective document rationale enrichments than existing augmentation baselines. An ablation study on RL optimization is provided in Appendix~\ref{app-RL-Optimization}.
\begin{table*}[t!]
\centering
\small
\setlength{\tabcolsep}{2pt}
\begin{tabular}{ll|ccccc|cccc|ccc|cc}
\toprule
\multicolumn{2}{c|}{\multirow{2.5}{*}{\textbf{Inference/Train Retriever}}}
& \multicolumn{5}{c|}{\textbf{Natural Language}}
& \multicolumn{4}{c|}{\textbf{Code}}
& \multicolumn{3}{c|}{\textbf{Math}}
& \multirow{2.5}{*}{\textbf{Avg.}} & \multirow{2.5}{*}{\textbf{Improv.}} \\
\cmidrule(r){3-7}\cmidrule(r){8-11}\cmidrule(r){12-14}
&&
\textbf{Bio.} & \textbf{Earth.} & \textbf{Econ.} & \textbf{Psy.} & \textbf{Sus.}
& \textbf{Rob.} & \textbf{Stack.} & \textbf{Leet.} & \textbf{Pony}
& \textbf{Aops} & \textbf{TheoQ.} & \textbf{TheoT.}
&& \\
\midrule

\multirow{4}{*}{\rotatebox[origin=c]{90}{\textbf{BGE}}}
& Baseline
& 11.7 & 24.4 & 16.4 & 17.4 & 13.1
& 11.7 & 10.6 & 26.7 & 5.7
& \textbf{6.0} & 13.0 & 6.9
& 13.6 & -- \\

& \cellcolor{gray!30}+RL-Index (BGE)
& \cellcolor{gray!30}\textbf{14.1}
& \cellcolor{gray!30}\textbf{27.2}
& \cellcolor{gray!30}16.9
& \cellcolor{gray!30}\textbf{18.9}
& \cellcolor{gray!30}14.0
& \cellcolor{gray!30}\textbf{14.0}
& \cellcolor{gray!30}14.0
& \cellcolor{gray!30}26.0
& \cellcolor{gray!30}10.5
& \cellcolor{gray!30}5.9
& \cellcolor{gray!30}\textbf{13.9}
& \cellcolor{gray!30}\textbf{9.6}
& \cellcolor{gray!30}\textbf{15.4}
& \cellcolor{gray!30}\textbf{+13.2\%} \\

& +RL-Index (SBERT)
& 13.2 & 26.8 & 16.5 & 18.8 & \textbf{14.1}
& 13.4 & \textbf{14.2} & \textbf{27.6} & 10.3
& 5.8 & 13.7 & 7.9
& 15.2 & +11.8\% \\

& +RL-Index (Qwen)
& 14.0 & 26.1 & \textbf{17.9} & 18.5 & 13.7
& 13.4 & 13.3 & 25.7 & \textbf{11.6}
& 4.6 & 13.5 & 8.3
& 15.1 & +11.0\% \\

\midrule

\multirow{4}{*}{\rotatebox[origin=c]{90}{\textbf{SBERT}}}
& Baseline
& 15.2 & 20.4 & 16.6 & \textbf{22.7} & 15.3
& 8.2 & 11.0 & 26.4 & 7.0
& 5.3 & 20.0 & 10.8
& 14.9 & -- \\

& +RL-Index (BGE)
& \textbf{17.9} & 20.8 & \textbf{19.5} & 21.6 & 15.1
& \textbf{11.1} & 13.7 & 27.0 & 6.0
& 5.0 & 19.9 & \textbf{15.2}
& 16.1 & +8.1\% \\

& \cellcolor{gray!30}+RL-Index (SBERT)
& \cellcolor{gray!30}15.7
& \cellcolor{gray!30}\textbf{22.5}
& \cellcolor{gray!30}18.9
& \cellcolor{gray!30}21.5
& \cellcolor{gray!30}16.1
& \cellcolor{gray!30}10.5
& \cellcolor{gray!30}\textbf{14.7}
& \cellcolor{gray!30}\textbf{28.3}
& \cellcolor{gray!30}\textbf{8.5}
& \cellcolor{gray!30}\textbf{5.4}
& \cellcolor{gray!30}\textbf{20.9}
& \cellcolor{gray!30}13.1
& \cellcolor{gray!30}\textbf{16.3}
& \cellcolor{gray!30}\textbf{+9.4\%} \\

& +RL-Index (Qwen)
& 15.4 & 23.0 & 18.5 & 21.6 & \textbf{16.3}
& 9.4 & 12.8 & 27.4 & 7.2
& 4.9 & 18.2 & 12.8
& 15.6 & +4.7\% \\

\midrule

\multirow{4}{*}{\rotatebox[origin=c]{90}{\textbf{Qwen}}}
& Baseline
& 29.9 & 39.6 & 17.7 & 24.4 & 20.3
& 13.2 & 21.2 & 25.5 & 12.4
& 14.4 & 27.8 & 32.9
& 23.3 & -- \\

& +RL-Index (BGE)
& \textbf{32.3} & 41.1 & \textbf{23.0} & 27.6 & 23.1
& 13.3 & 21.1 & 24.6 & 5.7
& 11.8 & \textbf{29.6} & 31.6
& 23.7 & +1.7\% \\

& +RL-Index (SBERT)
& 31.1 & \textbf{43.0} & 22.8 & 27.5 & 24.1
& 14.6 & 19.5 & 26.3 & 14.8
& 8.3 & 29.2 & 29.8
& 24.3 & +4.3\% \\

& \cellcolor{gray!30}+RL-Index (Qwen)
& \cellcolor{gray!30}29.8
& \cellcolor{gray!30}39.7
& \cellcolor{gray!30}21.9
& \cellcolor{gray!30}\textbf{27.8}
& \cellcolor{gray!30}\textbf{26.7}
& \cellcolor{gray!30}\textbf{16.6}
& \cellcolor{gray!30}\textbf{22.1}
& \cellcolor{gray!30}\textbf{28.3}
& \cellcolor{gray!30}\textbf{17.0}
& \cellcolor{gray!30}\textbf{16.0}
& \cellcolor{gray!30}28.5
& \cellcolor{gray!30}\textbf{33.6}
& \cellcolor{gray!30}\textbf{25.7}
& \cellcolor{gray!30}\textbf{+10.3\%} \\

\bottomrule
\end{tabular}
\vspace{-2ex}
\caption{Transferability of RL-Index across retrievers. Each block corresponds to the deployment (inference-time) retriever, while sub-rows indicate the retriever used to train the RL augmentor. Shaded rows denote matched training–deployment settings. RL-Index consistently improves performance under cross-retriever transfer, with the strongest results achieved when training and deployment retrievers are aligned.}
\label{tab-transfer-retriever}
\vspace{-3ex}
\end{table*}

\subsection{Transferability Analysis}
Motivated by the performance gains of rationale-augmented documents, we study their transferability. 
For retrievers, augmented documents may be fetched by inference-time retrievers that differ from the training-time retriever used for reward computation, raising the question of cross-retriever transferability. 
%
%
%
To evaluate retriever transferability, we fix LLaMA as the base rationale generator and train the augmentor with rewards derived from a dense retriever, then assess retrieval performance using a different retriever on the same augmented corpus. This setup tests whether the learned rationales are retriever-agnostic rather than overfitting to a specific embedding model. We report nDCG@10 on BRIGHT under this cross-retriever setting.
Table~\ref{tab-transfer-retriever} shows that the augmented documents consistently yield performance gains even when evaluated with a different retriever, indicating that the learned rationales generalize effectively across retriever architectures. 
This transferability stems from capturing latent logic gaps via natural language rationales, which yields a universal semantic signal beyond specific retrieval embedding spaces. Transferability across LLM-powered rationale augmentors can be found in Appendix~\ref{app-transfer}


\subsection{Question-answering Performance}
We examine whether improved retrieval from RL-Index can further enhance downstream QA performance, rather than extract query keywords as a shortcut. Following~\citet{lee2025imagine,hongjinbright}, we evaluate responses with GPT-4o as a judge by referring to BRIGHT gold answers.
Using SBERT, we provide Top-10 documents retrieved from offline (\textsc{SPIKE}, RL-Index), online (TongSearch) reasoning, and their combination, as context to the generators (Claude-Sonnet-4.5, Llama3.3-70B-Instruct, GPT-5). 

\begin{table}[t!]
\centering
\scriptsize
\setlength{\tabcolsep}{2pt}
\begin{tabular}{ll|ccccccc|c}
\toprule
\multirow{2.5}{*}{\textbf{Gen.}} & \multirow{2.5}{*}{\textbf{Method}}
& \multicolumn{7}{c|}{\textbf{Dataset}} 
& \multirow{2}{*}{\textbf{Avg.}} \\
\cmidrule(r){3-9}
& & \textbf{Bio.} & \textbf{Econ.} & \textbf{Psy.} & \textbf{Earth.} & \textbf{Stack.} & \textbf{Rob.} & \textbf{Sus.} & \\
\midrule

\multirow{5}{*}{\rotatebox[origin=c]{90}{\textbf{Claude}}}
& SBERT & 60.2 & 61.8 & 62.8 & 68.4 & 72.6 & 57.6 & 54.1 & 62.5 \\
& +SPIKE & 65.4 & 59.8 & 64.7 & 69.9 & 70.7 & 56.3 & 57.0 & 63.4 \\
& \cellgq+RL-Index  & \cellgq61.6 & \cellgq64.2 & \cellgq63.0 & \cellgq69.5 & \cellgq73.8 & \cellgq60.2 & \cellgq59.7 & \cellgq64.6 \\
& +TongSearch & 68.7 & 64.6 & 69.0 & 71.6 & 75.2 & 59.1 & 58.0 & 66.6 \\
& \cellcolor{gray!30}+TS\&RL-Index & \cellcolor{gray!30}68.6 & \cellcolor{gray!30}63.2 & \cellcolor{gray!30}69.2 & \cellcolor{gray!30}73.1 & \cellcolor{gray!30}74.4 & \cellcolor{gray!30}61.4 & \cellcolor{gray!30}61.6 & \cellcolor{gray!30}67.4\\

\midrule

\multirow{5}{*}{\rotatebox[origin=c]{90}{\textbf{Llama}}}
& SBERT & 56.5 & 53.3 & 58.0 & 61.8 & 63.4 & 44.5 & 53.1 & 55.8 \\
& +SPIKE & 59.6 & 55.1 & 56.4 & 55.7 & 62.6 & 43.0 & 51.7 & 54.9 \\
& \cellgq+RL-Index  & \cellgq59.1 & \cellgq55.4 & \cellgq62.0 & \cellgq60.7 & \cellgq61.0 & \cellgq42.7 & \cellgq55.5 & \cellgq56.6 \\
& +TongSearch &63.6 &59.6 &64.7 &65.3 &65.3 &53.4 &61.2 &61.9\\
& \cellgl+TS\&RL-Index & \cellgl65.0 & \cellgl57.6 & \cellgl62.8 & \cellgl65.0 & \cellgl67.1 & \cellgl55.1 & \cellgl61.3 & \cellgl62.0\\

\midrule
\multirow{5}{*}{\rotatebox[origin=c]{90}{\textbf{GPT}}}
&SBERT &69.1 &71.7 &70.8 &76.3 &75.0 &68.3 &69.2 &71.5 \\
&+SPIKE &72.0, &71.9 &71.2, &76.7 &76.3 &70.4 &70.4 &72.7\\
&\cellgq+RL-Index & \cellgq 70.3 &\cellgq 74.2 & \cellgq75.0 & \cellgq76.1 & \cellgq83.0 & \cellgq73.1 & \cellgq69.4 & \cellgq74.4 \\
&TongSearch & 70.8 & 72.2 & 73.3 & 77.1 & 76.6 & 68.6 & 66.2 &72.1\\
&\cellgl TS\&RL-Index & \cellgl 72.2 & \cellgl 72.4 & \cellgl 73.1 & \cellgl 81.0 & \cellgl 80.9 & \cellgl 71.7& \cellgl 71.9 & \cellgl 74.7\\

\bottomrule
\end{tabular}
\caption{QA performance when using retrieval contexts enhanced with reasoning-augmented documents and queries. Results demonstrate complementary gains from query/document-level reasoning.}
\label{tab-qa}
\vspace{-4ex}
\end{table}

In Table~\ref{tab-qa}, RL-Index consistently outperforms both the baseline and \textsc{SPIKE} across all three generators, showing that our offline augmentation method improves QA by enhancing retrieval relevance and providing richer context.
We further evaluate TongSearch and its combination with RL-Index (TS\&RL-Index), where query rewriting and document augmentation are jointly applied by using TongSearch's reasoned query and providing RL-Index augmented documents. TS\&RL-Index consistently surpasses TongSearch alone, indicating that RL-Index provides complementary document-level signals to query-side reasoning, leading to stronger end-to-end QA performance.

\begin{table*}[t]
\centering
\label{tab:efficiency_comparison}
\setlength{\tabcolsep}{4.2pt}
\resizebox{\textwidth}{!}{%
\begin{tabular}{l|c|c|c|c|c}
\hline
\textbf{Method} & \textbf{Query Reasoning (ms)} & \textbf{Query Embedding (ms)} & \textbf{Retrieval (ms)} & \textbf{Total (ms)} & \textbf{nDCG@10} \\ \hline
BGE & 0.0 &13.8 &55.8  &69.6  &13.6\\
+TongSearch & 7660.0 &21.1 &55.8 &7736.9  & 17.5\\
\cellgq+RL-Index & \cellgq0.0 &\cellgq13.8 &\cellgq100.8 &\cellgq114.6  & \cellgq15.4 \\ 
\cellgl+TS\&RL-Index &\cellgl7660.0 &\cellgl21.1 &\cellgl100.8 &\cellgl7781.9 &\cellgl19.3 \\\hline
SBERT & 0.0 & 9.7 & 45.0 & 54.7 &14.9\\
+TongSearch & 7660.0 & 11.0 & 45.0 & 7716.0 & 16.8\\
\cellgq+RL-Index & \cellgq0.0 & \cellgq9.7 & \cellgq69.8 & \cellgq79.5 & \cellgq16.3 \\ 
\cellgl+TS\&RL-Index &\cellgl7660.0 &\cellgl11.0 &\cellgl69.8 &\cellgl7740.8 &\cellgl18.1 \\\hline
%
\end{tabular}
}
\caption{Comparing online efficiency–effectiveness trade-offs, TongSearch improves nDCG@10 but incurs substantial query-time overhead, while RL-Index shifts reasoning offline, achieving comparable performance with much lower latency. Combining both yields peak effectiveness with latency nearly identical to TongSearch.}
\label{tab-online}
\vspace{-3ex}
\end{table*}
\subsection{Overall Retrieval Efficiency}\label{sec-efficiency}
\vspace{-1ex}
Since latency is a key factor affecting user satisfaction in retrieval, this section evaluates the efficiency of RL-Index against both online query reasoning and offline document reasoning baselines. By shifting reasoning from query-time rewriting to offline document augmentation, RL-Index eliminates online reasoning overhead. We evaluate (1) the online performance–latency trade-off and (2) offline token efficiency. Detailed results are in Appendix~\ref{app-online-latency}-\ref{app-offline-latency}.

\textbf{Online Performance–Latency Trade-off.} To evaluate online retrieval efficiency, we compare the per-query retrieval latency of our offline rationale augmentation RL-Index with the online query rewriting baseline TongSearch using SBERT and BGE. Table~\ref{tab-online} reveals a clear efficiency–effectiveness trade-off. Compared with a vanilla retriever without any reasoning augmentation, our RL-Index improves retrieval effectiveness with a modest latency increase, while remaining much faster than TongSearch. For BGE, RL-Index improves nDCG@10 from 13.6 to 15.4 (+13.2\%) with total latency increasing from 69.6 ms to 114.6 ms. Compared with TongSearch (7736.9 ms, 17.5 nDCG@10), RL-Index is about $68\times$ faster. For SBERT, RL-Index improves nDCG@10 from 14.9 to 16.3 (+9.4\%) at 79.5 ms total latency, whereas TongSearch reaches 16.8 at 7716.0 ms, making RL-Index about $97\times$ faster. This gap is primarily due to removing online query-rewriting reasoning time (0.0 ms for RL-Index vs. 7660.0 ms for TongSearch). Although the RL-Index slightly increases retrieval time due to the augmented index, overall latency remains much lower. Combining TongSearch with RL-Index can further achieve the best nDCG@10 (19.3 on BGE, 18.1 on SBERT) with latency nearly identical to TongSearch (7781.9/7740.8 ms). It suggests that document-level RL-Index reasoning effectively complements query-side TongSearch reasoning.

\begin{table}
\centering
\small
\setlength{\tabcolsep}{5.5pt}
\begin{tabular}{lcc}
\toprule
\textbf{Metric} & \textsc{SPIKE} & \textbf{RL-Index} \\
\midrule
Training API Tokens/Doc & 1,014 & \textbf{0} \\
Inference Tokens/Doc & 345.6 & \textbf{257.0} \\
Indexing Overhead (\# Aug Docs) & 387,391 & \textbf{111,097} \\
\bottomrule
\end{tabular}

\caption{Offline efficiency comparison.}
\label{tab-offline}
\end{table}

\textbf{Offline Indexing Token Efficiency.}
We compare RL-Index with \textsc{SPIKE} across three efficiency dimensions (Table~\ref{tab-offline}), including training cost, augmentation overhead, and indexing footprint. 
For training API Tokens/Doc, \textsc{SPIKE} relies on a GPT-4o-based distillation process to construct training data to optimize a smaller LLM as a document augmentor. Specifically, each document requires an API call consisting of 523 input tokens and 491 output tokens, resulting in a total of 1,014 API tokens. In contrast, RL-Index directly optimizes the augmentation model through reinforcement learning without requiring external LLM-generated supervision, incurring zero API-token cost during training.
We report generated tokens per document as a proxy to reflect inference efficiency. RL-Index generates fewer tokens per document during augmentation compared to SPIKE (257.0 vs. 345.6). Consequently, RL-Index requires less decoding computation when augmenting the large-scale document corpus while producing more compact but effective augmentations.
For Indexing overhead (\# Aug Docs), RL-Index also significantly reduces the augmented corpus, producing only 111,097 documents compared to 387,391 documents generated by SPIKE. The primary reason is that RL-Index adopts a one-to-one augmentation design where each original document only has one augmented document. In contrast, \textsc{SPIKE} creates multiple scenario-specific variants for each document.
Overall, RL-Index consistently reduces cost across training, augmentation, and indexing, resulting in a more efficient offline pipeline.



\begin{figure*}[htbp!]
\vspace{-2ex}
    \centering
    \includegraphics[width=1\linewidth]{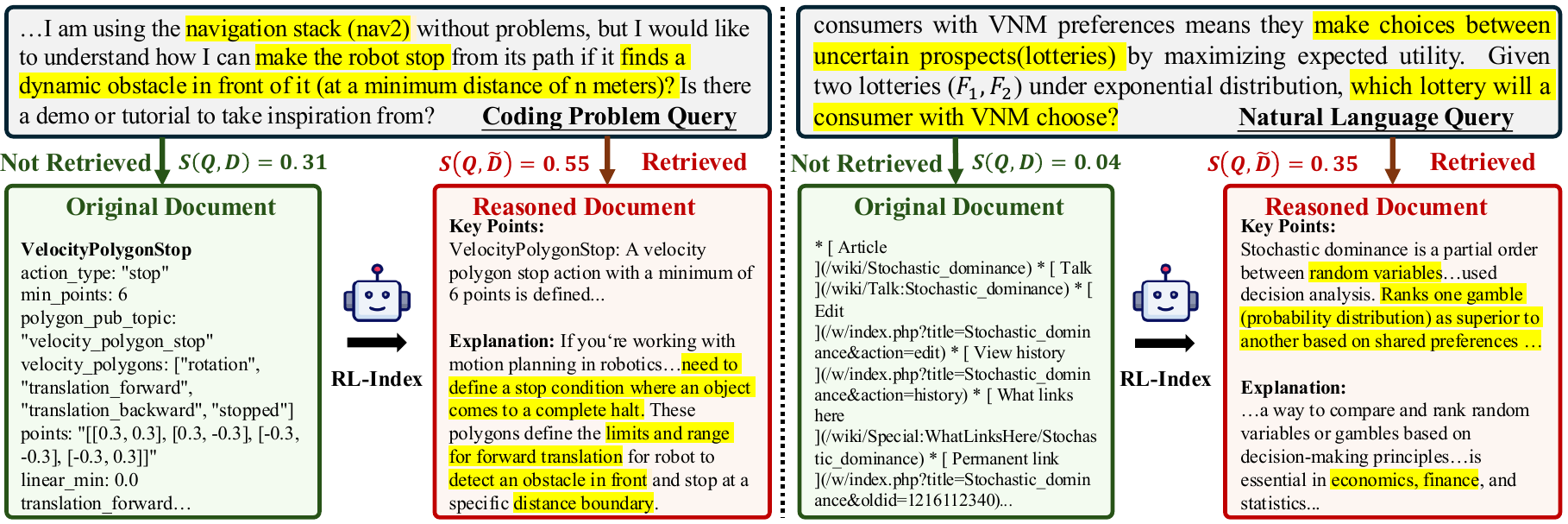}
    \vspace{-3ex}
    \caption{Case studies where RL-Index generates effective rationales for successful document retrieval.}
    \label{fig-Example}
    \vspace{-3ex}
\end{figure*}

\begin{table}[t!]
\centering
\scriptsize
\setlength{\tabcolsep}{2pt}
\begin{tabular}{lccccccc}
\toprule
\textbf{Metric} & \textbf{Bio.} & \textbf{Earth.} & \textbf{Econ.} & \textbf{Psy.} & \textbf{Sus.} & \textbf{Rob.} & \textbf{Stack.} \\
\midrule
\textbf{Ori Diff} &
0.024 & 0.028 & 0.023 & 0.031 & 0.019 & 0.011 & 0.005 \\

\textbf{Aug Diff} &
0.032 & 0.062 & 0.046 & 0.039 & 0.031 & 0.047 & 0.054 \\

\midrule
\textbf{Increase} &
+33.3\% & +121.4\% & +100.0\% & +25.8\% & +63.2\% & +327.3\% & +980.0\% \\
\bottomrule
\end{tabular}
\vspace{-1ex}
\caption{Comparison of query–document similarity margins before and after document augmentation. Original Diff and Augmented Diff are defined as $\mathrm{sim}(Q,D^{+})-\mathrm{sim}(Q,D^{-})$ and $\mathrm{sim}(Q,\tilde{D}^{+})-\mathrm{sim}(Q,\tilde{D}^{-})$, respectively. Document augmentation with generated rationales consistently increases the similarity margin, improving retrieval discriminability.}
\label{tab-margin_comparison}
\vspace{-5ex}
\end{table}

\vspace{-2ex}
\section{Rationale Generation Analysis}
\vspace{-2ex}
This section presents both quantitative and qualitative analyses to validate our approach. First, we evaluate our reward design through a statistical analysis of changes in the query–positive versus query–negative similarity margins (Table~\ref{tab-margin_comparison}). Second, we examine the quality of the generated rationales through a qualitative case study of the rationale content injected into the documents (Figure~\ref{fig-Example}), assessing whether they capture logical reasoning rather than merely retriever-preferred wording.

\textbf{Quantitative Analysis of Retrieval Similarity.} To justify that the query-positive-only reward used in RL training in Section~\ref{sec-rl} is sufficient, we analyze the similarity margin between positive and negative documents before and after augmentation. Specifically, for each query, we retrieve the top-10 documents using SBERT, consisting of the relevant document and hard negatives. We then compute the average query similarity to the positive and negative document sets. In Table~\ref{tab-margin_comparison}, the similarity margin between the query and positive versus hard-negative documents consistently increases across all domains after augmentation, with relative improvements ranging from 25.8\% to 980.0\%. These results indicate that optimizing only the query--positive similarity is sufficient to enhance retrieval discriminability without explicitly incorporating negative documents into the reward.

\textbf{Qualitative Analysis of Generated Rationale.} To understand whether RL-Index improves retrieval by generating the underlying logic, we present case studies from both natural language and code domains in Figure~\ref{fig-Example}.
In the code example, the original document includes the correct Nav2 settings, but it uses low-level configuration text that does not match the user’s natural-language intent. RL-Index adds an intent-based explanation (linking ``stop at a specific distance'' to polygon-based stop logic), resulting in a similarity increase from 0.31 to 0.55.
In the natural-language example, the original document is relevant but not retrieved as its wording focuses on providing links. RL-Index rewrites the document into clearer, query-aligned reasoning text, which raises similarity from 0.04 to 0.35 and makes retrieval succeed. More detailed retrieval case study analysis is in Appendix~\ref{app-case-study-retrieval} and further QA case study is in Appendix~\ref{app-case-study-qa}.

\section{Conclusion}
In this paper, we study reasoning-intensive document retrieval from an indexing perspective. Instead of relying on online query reasoning, we propose a reinforcement learning-based offline indexing framework that trains an LLM as a document augmenter to make latent document rationales explicit and facilitate retrieval. 
Our method combines rationale generation with GRPO-based policy optimization, and performs retrieval over both original and augmented documents. 
Extensive experiments show consistent gains in both retrieval and QA performance across diverse retrievers and generators. Compared to query-side reasoning, our framework shifts computation offline, achieving a better latency-accuracy trade-off at inference. 
%
In future work, we plan to explore diversity-aware rationale augmentation to provide complementary coverage of reasoning needs.

\newpage
\section{Limitations}
Although our RL-Index achieves superior performance in both retrieval and efficiency, it has two limitations. Firstly, we only use positive data during training. Our RL reward optimizes positive-pair similarity gain. Although it is simple but effective, it does not consider negative documents, making it a weak proxy for retrieval ranking. In the future, we will consider including negative documents during training with a modified reward design that also includes negative similarity. Secondly, existing work reasons either over the query or the document to reduce the semantic mismatch between user intent and document content. The two strategies have complementary advantages, where query reasoning enriches the representation of information needs while document reasoning explicates the knowledge provided. Since both optimize the same retrieval objective, training the two pipelines separately missed the opportunity to align their representations. Future work would explore using the unified framework to train both document reasoning and query reasoning simultaneously.

\bibliography{custom}

\newpage

\section{Dataset Details}\label{app-dataset}
BRIGHT comprises 1,398 real-world queries spanning diverse domains, including economics, psychology, robotics, mathematics, and software programming. These queries are designed to reflect challenging scenarios that require deep comprehension and reasoning to retrieve relevant documents. BRIGHT is categorized~\cite{hongjinbright} into three groups: StackExchange, Coding, and Theorem-based collections, which are:
\begin{itemize}[leftmargin=*]
    \item \textbf{StackExchange}: Biology (\textbf{Bio.}), Earth Science (\textbf{Earth.}), Economics (\textbf{Econ.}), Psychology (\textbf{Psy.}), Robotics (\textbf{Rob.}), Stack Overflow (\textbf{Stack.}), Sustainable Living (\textbf{Sus.})
    \item \textbf{Coding}: Leetcode (\textbf{Leet.}), Pony (\textbf{Pony})
    \item \textbf{Theorem}: AoPS (\textbf{AoPS}), TheoremQA-Question (\textbf{TheoQ.}), TheoremQA-Theorem (\textbf{TheoT.})
\end{itemize}

We follow~\citet{lee2025imagine} and adopt a classification based on document type to better capture retrieval challenges arising from different content structures. Specifically, we group the datasets into Natural Language, Code, and Math as follows:
\begin{itemize}[leftmargin=*]
    \item \textbf{Natural Language}: Biology (\textbf{Bio.}), Earth Science (\textbf{Earth.}), Economics (\textbf{Econ.}), Psychology (\textbf{Psy.}), Sustainable Living (\textbf{Sus.})
    \item \textbf{Code}: Leetcode (\textbf{Leet.}), Pony (\textbf{Pony}), Robotics (\textbf{Rob.}), Stack Overflow (\textbf{Stack.})
    \item \textbf{Math}: AoPS (\textbf{AoPS}), TheoremQA-Question (\textbf{TheoQ.}), TheoremQA-Theorem (\textbf{TheoT.})
\end{itemize}

\section{Experimental Outcomes}
\subsection{Transferability across LLM-powered Rationale Generator.}\label{app-transfer}
To evaluate LLM transferability, we replace the base rationale augmentor with Qwen and repeat the same training/evaluation pipeline. This experiment examines whether our RL-based rationale augmentation generalizes across different LLM families.
In Table~\ref{tab-transfer-llm}, Qwen-based augmentation exhibits similar improvement trends to the LLaMA-based setting shown in Table~\ref{tab-unified-llama} before across most tasks. This indicates that the GRPO training objective and rationale format are not tied to a specific type of LLM, and the performance gains over other query and document reasoning baselines remain when switching the underlying augmentors.

\begin{table*}[htbp!]
\centering
\small
\setlength{\tabcolsep}{1.5pt}
\begin{tabular}{c|c|ccccc|cccc|ccc|cc}
\toprule
\multirow{2.5}{*}{\textbf{Retriever}} & \multirow{2.5}{*}{\makecell[c]{\textbf{Rationale}\\\textbf{Augmentor}}}
& \multicolumn{5}{c|}{\textbf{Natural Language}} 
& \multicolumn{4}{c|}{\textbf{Code}} 
& \multicolumn{3}{c|}{\textbf{Math}} 
& \multirow{2}{*}{\textbf{Avg.}} & \multirow{2}{*}{\textbf{Improv.}} \\
\cmidrule(r){3-14}
& & \textbf{Bio.} & \textbf{Earth.} & \textbf{Econ.} & \textbf{Psy.} & \textbf{Sus.}
& \textbf{Rob.} & \textbf{Stack.} & \textbf{Leet.} & \textbf{Pony}
& \textbf{Aops} & \textbf{TheoQ.} & \textbf{TheoT.}
&  &  \\
\midrule

\multirow{3}{*}{\rotatebox{90}{\textbf{BGE}}}
& Baseline
& 11.7 &24.4 &16.4 &17.4 &13.1 &11.7 &10.6 &26.7 &5.7 &6.0 &13.0 &6.9 &13.6 &-- \\

& \cellgq +RL-Index (Qwen)
& \cellgq 12.0 &\cellgq 26.2 &\cellgq 16.1 &\cellgq 17.9 &\cellgq 13.8
& \cellgq 12.2 &\cellgq 12.0 &\cellgq 26.8 &\cellgq 10.5
& \cellgq 5.8 &\cellgq 13.2 &\cellgq 7.3
& \cellgq 14.5 &\cellgq +6.6\% \\

& \cellgl +RL-Index (Llama)
& \cellgl 14.1 &\cellgl 27.2 &\cellgl 16.9 &\cellgl 18.9 &\cellgl 14.0
& \cellgl 14.0 &\cellgl 14.0 &\cellgl 26.0 &\cellgl 10.5
& \cellgl 5.9 &\cellgl 13.9 &\cellgl 9.6
& \cellgl 15.4 &\cellgl +13.2\% \\

\hline

\multirow{3}{*}{\rotatebox{90}{\textbf{SBERT}}}
& Baseline
& 15.2 &20.4 &16.6 &22.7 &15.3 &8.2 &11.0 &26.4 &7.0 &5.3 &20.0 &10.8 &14.9 &-- \\

& \cellgq +RL-Index (Qwen)
& \cellgq 15.5 &\cellgq 22.3 &\cellgq 17.4 &\cellgq 21.3 &\cellgq 15.8
& \cellgq 9.0 &\cellgq 12.0 &\cellgq 27.1 &\cellgq 7.1
& \cellgq 5.8 &\cellgq 21.0 &\cellgq 13.2
& \cellgq 15.6 &\cellgq +4.7\% \\

& \cellgl +RL-Index (Llama)
& \cellgl 15.7 &\cellgl 22.5 &\cellgl 18.9 &\cellgl 21.5 &\cellgl 16.1
& \cellgl 10.5 &\cellgl 14.7 &\cellgl 28.3 &\cellgl 8.5
& \cellgl 5.4 &\cellgl 20.9 &\cellgl 13.1
& \cellgl 16.3 &\cellgl +9.4\% \\

\hline

\multirow{3}{*}{\rotatebox{90}{\textbf{Qwen}}}
& Baseline
&29.9 &39.6 &17.7 &24.4 &20.3 &13.2 &21.2 &25.5 &12.4 &14.4 &27.8 &32.9 &23.3 &-- \\

& \cellgq +RL-Index (Qwen)
& \cellgq 30.4 &\cellgq 40.0 &\cellgq 20.4 &\cellgq 25.4 &\cellgq 21.4
& \cellgq 12.9 &\cellgq 22.2 &\cellgq 26.1 &\cellgq 17.6
& \cellgq 15.3 &\cellgq 27.6 &\cellgq 34.7
& \cellgq 24.5 &\cellgq +5.2\% \\

& \cellgl +RL-Index (Llama)
& \cellgl 29.8 &\cellgl 39.7 &\cellgl 21.9 &\cellgl 27.8 &\cellgl 26.7
& \cellgl 16.6 &\cellgl 22.1 &\cellgl 28.3 &\cellgl 17.0
& \cellgl 16.0 &\cellgl 28.5 &\cellgl 33.6
& \cellgl 25.7 &\cellgl +10.3\% \\
\bottomrule
\end{tabular}
\vspace{-1.5ex}
\caption{Comparison of RL-Index trained with different LLM-based rationale augmentors. Results show that the RL-Index pipeline remains effective across different LLM-based agents, demonstrating its robustness to the choice of underlying augmentor. Gray and darker gray cells denote Qwen and LLaMA augmentations, respectively.}
\label{tab-transfer-llm}
\vspace{-3ex}
\end{table*}

\begin{table}[t!]
\centering
\begin{tabular}{lrr}
\hline
\textbf{Metric} & \textbf{Train} & \textbf{Eval} \\
\hline
Documents & 7,875 & 84 \\
Total input tokens & 4,112,682 & 47,219 \\
Total output tokens & 3,867,870 & 43,399 \\
Avg input tokens/doc & 522.2 & 562.1 \\
Avg output tokens/doc & 491.2 & 516.7 \\
\hline
\end{tabular}
\caption{GPT-4o cost of SPIKE.}
\label{tab-spike-api}
\vspace{-4ex}
\end{table}

\vspace{-2ex}
\subsection{Online Latency Analysis}\label{app-online-latency}
\vspace{-1ex}
To evaluate the online efficiency of our method, we compare baseline retrievers (e.g., BGE, SBERT), +RL-Index (trained based on Llama3.2-3B-Instruct), the query reasoner +TongSearch (trained based on Qwen2.5-1.5B-Instruct), and the one combining two augmentation methods (i.e., +TS\&RL-Index). We first report per-query reasoning latency across datasets for TongSearch (Table~\ref{tab-qr-tongsearch}). While TongSearch takes around 7.66s to reason for one query, our RL-Index introduces zero online reasoning latency because reasoning is shifted entirely to the offline indexing stage. 
\begin{table*}[htbp!]
\centering
\small
\setlength{\tabcolsep}{4.7pt}
\begin{tabular}{l|ccccc|cccc|ccc|c}
\toprule
\multirow{2.5}{*}{\textbf{Metric}}
& \multicolumn{5}{c|}{\textbf{Natural Language}}
& \multicolumn{4}{c|}{\textbf{Code}}
& \multicolumn{3}{c|}{\textbf{Math}}
& \multirow{2.5}{*}{\textbf{Avg.}} \\
\cmidrule(r){2-6}\cmidrule(r){7-10}\cmidrule(r){11-13}
& \textbf{Bio.} & \textbf{Earth.} & \textbf{Econ.} & \textbf{Psy.} & \textbf{Sus.}
& \textbf{Rob.} & \textbf{Stack.} & \textbf{Leet.} & \textbf{Pony}
& \textbf{Aops} & \textbf{TheoQ.} & \textbf{TheoT.}
& \\
\midrule

\textbf{Num Queries} 
& 103 & 116 & 103 & 101 & 108
& 101 & 117 & 142 & 112
& 111 & 194 & 76
& 115.3 \\

\textbf{Avg Time (s)}
& 7.00 & 7.89 & 7.32 & 7.58 & 7.08
& 8.14 & 7.79 & 7.48 & 7.04
& 9.34 & 7.54 & 7.76
& 7.66 \\

\textbf{Total Est (h)}
& 0.20 & 0.25 & 0.21 & 0.21 & 0.21
& 0.23 & 0.25 & 0.30 & 0.22
& 0.29 & 0.41 & 0.16
& 0.25 \\

\bottomrule
\end{tabular}
\vspace{-2ex}
\caption{Average online query reasoning time across datasets using TongSearch, grouped by data type. Our proposed RL-Index introduces zero online query reasoning latency.}
\label{tab-qr-tongsearch}
\vspace{-2ex}
\end{table*}

We then measure latency in the online retrieval stage, which includes query embedding and index search. Detailed measurements across various datasets using different retrievers are reported in Table~\ref{tab-online-retrieval-bge} and Table~\ref{tab-online-retrieval-sbert}, and each value means the latency per query. Based on the results, we have three observations. First, since RL-Index avoids query reasoning, it usually encodes shorter inputs, resulting in lower query embedding latency compared with TongSearch across both retrievers (BGE: 0.0138s vs.
0.0211s; SBERT: 0.0097s vs.
0.0110s). 
Second, since RL-Index performs retrieval over an expanded index, it has higher retrieval latency (BGE: 0.1008s vs.
0.0558s; SBERT: 0.0698s vs.
0.0452s).
Third, despite the additional retrieval overhead, RL-Index remains competitive in effectiveness, achieving 15.4 nDCG@10 with BGE and 16.3 with SBERT. Importantly, it maintains substantially lower end-to-end online latency than query-side reasoning methods by eliminating query-time reasoning and rewriting, which accounts for a significant portion of latency (e.g., an average of 7.66s in Table~\ref{tab-qr-tongsearch}). Overall, these results validate the design choice of shifting reasoning from online query processing to offline indexing, enabling a better latency–quality trade-off.

\begin{table*}[t!]
\centering
\scriptsize
\setlength{\tabcolsep}{4.5pt}
\begin{tabular}{l|ccccc|cccc|ccc|c}
\toprule
\multirow{2.5}{*}{\textbf{Metric}}
& \multicolumn{5}{c|}{\textbf{Natural Language}}
& \multicolumn{4}{c|}{\textbf{Code}}
& \multicolumn{3}{c|}{\textbf{Math}}
& \multirow{2.5}{*}{\textbf{Avg.}} \\
\cmidrule(r){2-6}\cmidrule(r){7-10}\cmidrule(r){11-13}
& \textbf{Bio.} & \textbf{Earth.} & \textbf{Econ.} & \textbf{Psy.} & \textbf{Sus.}
& \textbf{Rob.} & \textbf{Stack.} & \textbf{Leet.} & \textbf{Pony}
& \textbf{Aops} & \textbf{TheoQ.} & \textbf{TheoT.}
& \\
\midrule

\textbf{Embedding (BGE)}
& 0.0158 & 0.0117 & 0.0129 & 0.0125 & 0.0125
& 0.0146 & 0.0184 & 0.0164 & 0.0118
& 0.0120 & 0.0115 & 0.0154
& 0.0138 \\

\textbf{Embedding (+TS)}
& 0.0213 & 0.0178 & 0.0206 & 0.0209 & 0.0239
& 0.0213 & 0.0205 & 0.0197 & 0.0254
& 0.0231 & 0.0206 & 0.0180
& 0.0211 \\

\textbf{Embedding (+RL)}
& 0.0158 & 0.0117 & 0.0129 & 0.0125 & 0.0125
& 0.0146 & 0.0184 & 0.0164 & 0.0118
& 0.0120 & 0.0115 & 0.0154
& 0.0138 \\

\textbf{Embedding (+TS\&RL)}
& 0.0213 & 0.0178 & 0.0206 & 0.0209 & 0.0239
& 0.0213 & 0.0205 & 0.0197 & 0.0254
& 0.0231 & 0.0206 & 0.0180
& 0.0211 \\

\midrule

\textbf{Retrieval (BGE)}
& 0.0268 & 0.0469 & 0.0245 & 0.0253 & 0.0305
& 0.0285 & 0.0432 & 0.1459 & 0.0186
& 0.1497 & 0.1141 & 0.0153
& 0.0558 \\

\textbf{Retrieval (+TS)}
& 0.0268 & 0.0469 & 0.0245 & 0.0253 & 0.0305
& 0.0285 & 0.0432 & 0.1459 & 0.0186
& 0.1497 & 0.1141 & 0.0153
& 0.0558 \\

\textbf{Retrieval (+RL)}
& 0.0537 & 0.0941 & 0.0485 & 0.0520 & 0.0555
& 0.0563 & 0.0852 & 0.2874 & 0.0172
& 0.2362 & 0.1940 & 0.0298
& 0.1008 \\

\textbf{Retrieval (+TS\&RL)}
& 0.0537 & 0.0941 & 0.0485 & 0.0520 & 0.0555
& 0.0563 & 0.0852 & 0.2874 & 0.0172
& 0.2362 & 0.1940 & 0.0298
& 0.1008 \\

\midrule
\textbf{nDCG@10 (BGE)} & 11.7 & 24.4 & 16.4 & 17.4 & 13.1 & 11.7 & 10.6 & 26.7 & 5.7 & 6.0 & 13.0 & 6.9 & 13.6 \\

\textbf{nDCG@10 (+TS)}
& 18.8 & 33.2 & 19.7 & 20.3 & 17.2
& 12.8 & 15.5 & 22.9 & 5.6
& 6.9 & 19.0 & 18.0
& 17.5 \\

\textbf{nDCG@10 (+RL)}
& 14.1 & 27.2 & 16.9 & 18.9 & 14.0
& 14.0 & 14.0 & 26.0 & 10.5
& 5.9 & 13.9 & 9.6
& 15.4 \\

\textbf{nDCG@10 (+TS\&RL)}
&20.6 & 33.5 & 19.9 & 21.5 & 16.7 & 15.1 & 17.7 &24.3 & 9.6 & 6.7 & 21.7 & 24.6 &19.3
 \\

\bottomrule
\end{tabular}
\vspace{-1ex}
\caption{Online retrieval latency and effectiveness between TongSearch (TS) and RL-Index (RL) using retriever BGE. Embedding time is the computation of query representations, whereas retrieval time captures similarity matching and candidate selection.}
\label{tab-online-retrieval-bge}
\vspace{-2ex}
\end{table*}

\begin{table*}[t!]
\centering
\scriptsize
\setlength{\tabcolsep}{4.5pt}
\begin{tabular}{l|ccccc|cccc|ccc|c}
\toprule
\multirow{2.5}{*}{\textbf{Metric}}
& \multicolumn{5}{c|}{\textbf{Natural Language}}
& \multicolumn{4}{c|}{\textbf{Code}}
& \multicolumn{3}{c|}{\textbf{Math}}
& \multirow{2.5}{*}{\textbf{Avg.}} \\
\cmidrule(r){2-6}\cmidrule(r){7-10}\cmidrule(r){11-13}
& \textbf{Bio.} & \textbf{Earth.} & \textbf{Econ.} & \textbf{Psy.} & \textbf{Sus.}
& \textbf{Rob.} & \textbf{Stack.} & \textbf{Leet.} & \textbf{Pony}
& \textbf{Aops} & \textbf{TheoQ.} & \textbf{TheoT.}
& \\
\midrule

\textbf{Embedding (SBERT)}
& 0.0095 & 0.0092 & 0.0097 & 0.0098 & 0.0097
& 0.0102 & 0.0101 & 0.0104 & 0.0094
& 0.0095 & 0.0088 & 0.0101
& 0.0097 \\

\textbf{Embedding (+TS)}
& 0.0112 & 0.0111 & 0.0112 & 0.0113 & 0.0115
& 0.0111 & 0.0110 & 0.0104 & 0.0109
& 0.0108 & 0.0105 & 0.0116
& 0.0110 \\

\textbf{Embedding (+RL)}
& 0.0095 & 0.0092 & 0.0097 & 0.0098 & 0.0097
& 0.0102 & 0.0101 & 0.0104 & 0.0094
& 0.0095 & 0.0088 & 0.0101
& 0.0097 \\

\textbf{Embedding (+TS\&RL)}
& 0.0112 & 0.0111 & 0.0112 & 0.0113 & 0.0115
& 0.0111 & 0.0110 & 0.0104 & 0.0109
& 0.0108 & 0.0105 & 0.0116
& 0.0110 \\

\midrule

\textbf{Retrieval (SBERT)}
& 0.0227 & 0.0382 & 0.0205 & 0.0211 & 0.0233
& 0.0239 & 0.0352 & 0.1125 & 0.0098
& 0.1270 & 0.0960 & 0.0124
& 0.0452 \\

\textbf{Retrieval (+TS)}
& 0.0227 & 0.0382 & 0.0205 & 0.0211 & 0.0233
& 0.0239 & 0.0352 & 0.1125 & 0.0098
& 0.1270 & 0.0960 & 0.0124
& 0.0452 \\

\textbf{Retrieval (+RL)}
& 0.0376 & 0.0683 & 0.0343 & 0.0358 & 0.0395
& 0.0405 & 0.0619 & 0.2104 & 0.0110
& 0.1494 & 0.1291 & 0.0195
& 0.0698 \\

\textbf{Retrieval (+TS\&RL)}
& 0.0376 & 0.0683 & 0.0343 & 0.0358 & 0.0395
& 0.0405 & 0.0619 & 0.2104 & 0.0110
& 0.1494 & 0.1291 & 0.0195
& 0.0698 \\

\midrule

\textbf{nDCG@10 (SBERT)} &15.2 & 20.4 & 16.6 & 22.7 & 15.3 & 8.2 & 11.0 & 26.4 & 7.0 & 5.3 & 20.0 & 10.8 & 14.9 \\

\textbf{nDCG@10 (+TS)}
& 17.9 & 24.2 & 18.5 & 24.5 & 15.0
& 9.7 & 12.8 & 17.9 & 25.2
& 6.1 & 6.6 & 22.6
& 16.8 \\

\textbf{nDCG@10 (+RL)}
& 15.7 & 22.5 & 18.9 & 21.5 & 16.1
& 10.5 & 14.7 & 13.1 & 28.3
& 8.5 & 5.4 & 20.9
& 16.3 \\

\textbf{nDCG@10 (+TS\&RL)} &16.7 &27.2 &20.7 &23.3 &16.3 &12.6 &14.8 &28.0 &5.8 &6.3 &22.8 &22.3 &18.1 \\

\bottomrule
\end{tabular}
\caption{Online retrieval latency and effectiveness between TongSearch (TS) and RL-Index (RL) using retriever SBERT. Embedding time is the computation of query representations, whereas retrieval time captures similarity matching and candidate selection.}
\label{tab-online-retrieval-sbert}
\end{table*}

\vspace{-2ex}
\subsection{Offline Latency Analysis}\label{app-offline-latency}
\vspace{-1ex}
To estimate the offline latency and indexing overhead of RL-Index, we compare our method with the offline reasoning framework, SPIKE. During training, SPIKE constructs scenario augmentations for supervised fine-tuning by prompting GPT-4o to reason over documents, which introduces additional external API cost in Table~\ref{tab-spike-api}. In contrast, RL-Index is trained with RL using only query--document pairs and does not require costly GPT-generated document augmentations for training.

After training, both methods augment the corpus and prepare the knowledge base before deploying the retrieval system for inference. To approximate document augmentation latency, we use the average number of generated tokens per document as a proxy for generation time. In Tables~\ref{tab-offline-token}, RL-Index consistently generates fewer tokens than SPIKE, indicating lower augmentation-time overhead while maintaining stronger retrieval effectiveness. In addition, generated augmentations will be encoded and stored in the knowledge base. This indexing overhead is approximated by the average number of augmentation documents that require embedding. Table~\ref {tab-offline-token} shows that SPIKE produces substantially more augmentation documents, as it decomposes multiple user scenarios into separate documents. Overall, RL-Index reduces both generation and indexing overhead while achieving better performance, resulting in a better offline pipeline.

\begin{table*}[htbp!]
\centering
\scriptsize
\vspace{-3ex}
\setlength{\tabcolsep}{3.5pt}
\begin{tabular}{l|ccccc|cccc|ccc|c}
\toprule
\multirow{2.5}{*}{\textbf{Metric}}
& \multicolumn{5}{c|}{\textbf{Natural Language}}
& \multicolumn{4}{c|}{\textbf{Code}}
& \multicolumn{3}{c|}{\textbf{Math}}
& \multirow{2.5}{*}{\textbf{Avg.}} \\
\cmidrule(r){2-6}\cmidrule(r){7-10}\cmidrule(r){11-13}
& \textbf{Bio.} & \textbf{Earth.} & \textbf{Econ.} & \textbf{Psy.} & \textbf{Sus.}
& \textbf{Rob.} & \textbf{Stack.} & \textbf{Leet.} & \textbf{Pony}
& \textbf{Aops} & \textbf{TheoQ.} & \textbf{TheoT.}
& \\
\midrule
\textbf{Tokens (SPIKE)}
& 288.5 & 189.2 & 225.6 & 279.9 & 312.1
& 267.8 & 378.0 & 432.5 & 346.7
& 473.4 & 490.0 & 463.3
& 345.6 \\

\textbf{Tokens (RL-Index)}
& 202.2 & 203.5 & 215.8 & 210.0 & 201.4
& 234.4 & 302.6 & 388.5 & 280.9
& 292.1 & 292.0 & 261.1
& 257.0 \\

\midrule

\textbf{\#Docs (SPIKE)}
& 146,257 & 206,414 & 96,777 & 148,864 & 180,439
& 159,520 & 387,179 & 1,720,933 & 26,155
& 724,960 & 766,663 & 84,532
& 387,391 \\

\textbf{\#Docs (RL-Index)}
& 57,359 & 121,249 & 50,220 & 52,835 & 60,792
& 61,961 & 107,081 & 413,932 & 7,894
& 188,002 & 188,002 & 23,839
& 111,097 \\

\bottomrule
\end{tabular}
\vspace{-1ex}
\caption{Comparison of offline augmentation cost between SPIKE and RL-Index.}
\label{tab-offline-token}
\vspace{-2ex}
\end{table*}

\vspace{-1ex}
\subsection{Ablation Study on RL Optimization}\label{app-RL-Optimization}
\vspace{-1ex}
\begin{table*}[t!]
\centering
\label{tab:model_performance}
\resizebox{\textwidth}{!}{%
\begin{tabular}{l *{14}{c}}
\toprule
Model & Bio. & Earth. & Econ. & Psy. & Sus. & Rob. & Stack. & Leet. & Pony & Aops & TheoQ. & TheoT. & Avg. & Improv. \\
\midrule
BGE & 11.7 & 24.4 & 16.4 & 17.4 & 13.1 & 11.7 & 10.6 & \textbf{26.7} & 5.7 & \textbf{6.0} & 13.0 & 6.9 & 13.6 & - \\
+RL-Index & \textbf{14.1} & \textbf{27.2} & \textbf{16.9} & \textbf{18.9} & \textbf{14.0} & \textbf{14.0} & \textbf{14.0} & 26.0 & \textbf{10.5} & 5.9 & \textbf{13.9} & \textbf{9.6} & \textbf{15.4} & \textbf{+13.2\%} \\
\textbf{+RL-Index W/O RL} & 10.9 & 23.9 & 16.6 & 16.4 & 13.5 & 12.6 & 12.7 & 26.1 & 5.6 & 5.2 & 13.0 & 5.1 & 13.5 & -0.74\% \\
\midrule
SBERT & 15.2 & 20.4 & 16.6 & \textbf{22.7} & 15.3 & 8.2 & 11.0 & 26.4 & 7.0 & 5.3 & 20.0 & 10.8 & 14.9 & -- \\
+RL-Index & \textbf{15.7} & \textbf{22.5} & \textbf{18.9} & 21.5 & \textbf{16.1} & 10.5 & \textbf{14.7} & \textbf{28.3} & 8.5 & \textbf{5.4} & \textbf{20.9} & 13.1 & \textbf{16.3} & \textbf{+9.4\%} \\
\textbf{RL-Index W/O RL} & 15.5 & 21.7 & 18.1 & 22.4 & 15.2 & \textbf{10.7} & 12.8 & 25.9 & \textbf{11.1} & 4.1 & 19.6 & \textbf{15.4} & 16.0 & +7.4\% \\
\midrule
Qwen & \textbf{29.9} & 39.6 & 17.7 & 24.4 & 20.3 & 13.2 & 21.2 & 25.5 & 12.4 & 14.4 & 27.8 & 32.9 & 23.3 & - \\
+RL-Index & 29.8 & \textbf{39.7} & \textbf{21.9} & \textbf{27.8} & \textbf{26.7} & \textbf{16.6} & \textbf{22.1} & \textbf{28.3} & \textbf{17.0} & \textbf{16.0} & \textbf{28.5} & \textbf{33.6} & \textbf{25.7} & \textbf{+10.3\%} \\
\textbf{+RL-Index W/O RL} & 17.7 & 30.5 & 19.2 & 24.4 & 17.5 & 10.8 & 16.6 & 23.4 & 11.9 & 3.0 & 20.0 & 10.8 & 17.2 & -26.2\% \\
\bottomrule
\end{tabular}%
}
\caption{Ablation study of RL optimization using the same LLM model (i.e., Llama-3.2-3B-Instruct) and the same prompt across various evaluation retrievers.}
\vspace{-2ex}
\label{tab-RL-ablation}
\end{table*}

To investigate the effectiveness of RL optimization, Table~\ref{tab-RL-ablation} compares our GRPO-optimized RL-Index against a prompt-only baseline (RL-Index W/O RL) using the same format~\ref{tab-prompt} and model (Llama-3.2-3B-Instruct). Relying solely on prompted rationales is insufficient: while it modestly improves SBERT (+7.4\%), it degrades average performance for BGE (-0.74\%) and Qwen (-26.2\%). In contrast, RL-Index consistently achieves the highest average performance across all three retrievers. This confirms that our performance gains do not come merely from adding a reasonable prompt instructing for rationale generation, but also from RL optimization successfully aligning rationale generation with actual retrieval preferences. Interpretable example analysis in Appendix~\ref{app-case-study-retrieval} further intuitively verifies the rationale generation.



\begin{figure}[t!]
    \centering
    \includegraphics[width=1\linewidth]{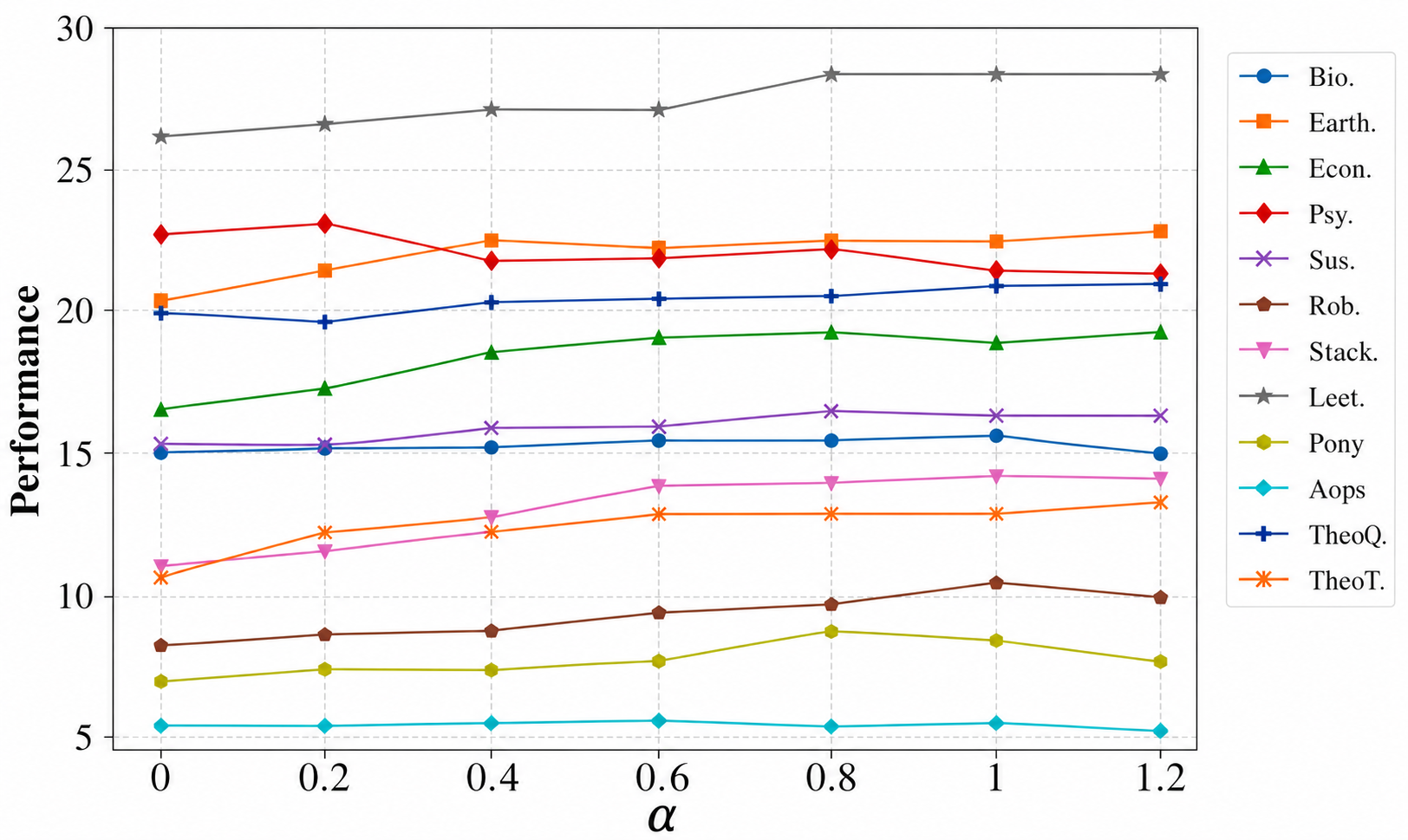}
    \caption{The performance change based on different combination weight $\alpha$ and across various domains.}
    \label{fig-sens}
\end{figure}

\subsection{Sensitivity Study of Combination Weight}~\label{app-alpha}
Since $\alpha$ is fixed to 1 throughout the paper, we further evaluate its sensitivity across all 12 domains by varying it from 0.0 to 1.2, as shown in Figure~\ref{fig-sens}. Overall, the results demonstrate that RL-Index is relatively robust to changes in $\alpha$. The average performance improves from 14.9 at $\alpha=0$ to 16.4 at $\alpha=0.8$, remains highly competitive at $\alpha=1$ (16.3), and only slightly declines at $\alpha=1.2$ (16.2). This suggests that the method is not highly sensitive to moderate variations in the combination weight and maintains competitive performance across a relatively broad range of $\alpha \in [0.8,1.2]$.
We further observe domain-specific variation in the optimal value of $\alpha$. For example, Biology (15.7), Robotics (10.5), and StackOverflow (14.7) peak at $\alpha = 1$, indicating that equally weighting the original document and augmented rationale is most effective. In contrast, Psychology (23.0) and Pony (8.9) achieve their best performance at $\alpha = 0.2$ and $\alpha = 0.8$, respectively, suggesting that their documents already possess stronger lexical or semantic alignment with user queries and therefore require less reliance on our proposed rationale augmentation. Overall, $\alpha=1$ provides a robust choice, despite the variation in domain-specific optimal values.
\newpage

\section{Case Study}
\subsection{Retrieval Example}\label{app-case-study-retrieval}
To understand why RL-Index improves retrieval over indexing only the original documents, we present case studies from both the natural language (Figure~\ref{fig-retrieval-nl}) and code (Figure~\ref{fig-retrieval-code}) domains, and show how the query and documents are related in the Reasoning Trace.
In the natural language example, the query asks how a consumer chooses between two lotteries under an exponential distribution. Although the ground-truth source document is relevant, it is not retrieved by the baseline because it mainly contains reference links and has weak lexical/semantic alignment with the query. In contrast, the RL-Index augmented document rewrites the same source content into more explicit and query-aligned reasoning statements, which substantially increases query-document similarity (from 0.04 to 0.35) and enables successful retrieval.
In the code-domain example, the user asks how to make a robot stop at a specific distance from a dynamic obstacle in Nav2. Although the original document contains the correct configuration (e.g., \texttt{VelocityPolygonStop}, \texttt{action\_type="stop"}, and polygon points), it is written in low-level configuration text and aligns poorly with the natural-language user query. As a result, the similarity between this query and the document is low ($s(q,d)=0.31$), so the relevant document is not retrieved. Instead, RL-Index rewrites the same content into an intent-oriented explanation that directly links ``stop at a specific distance'' to the polygon-based stop logic, increasing similarity from 0.31 to 0.55 and leading to successful retrieval.

\subsection{QA Example}\label{app-case-study-qa}
\begin{figure*}[t!]
    \centering
    \includegraphics[width=1\linewidth]{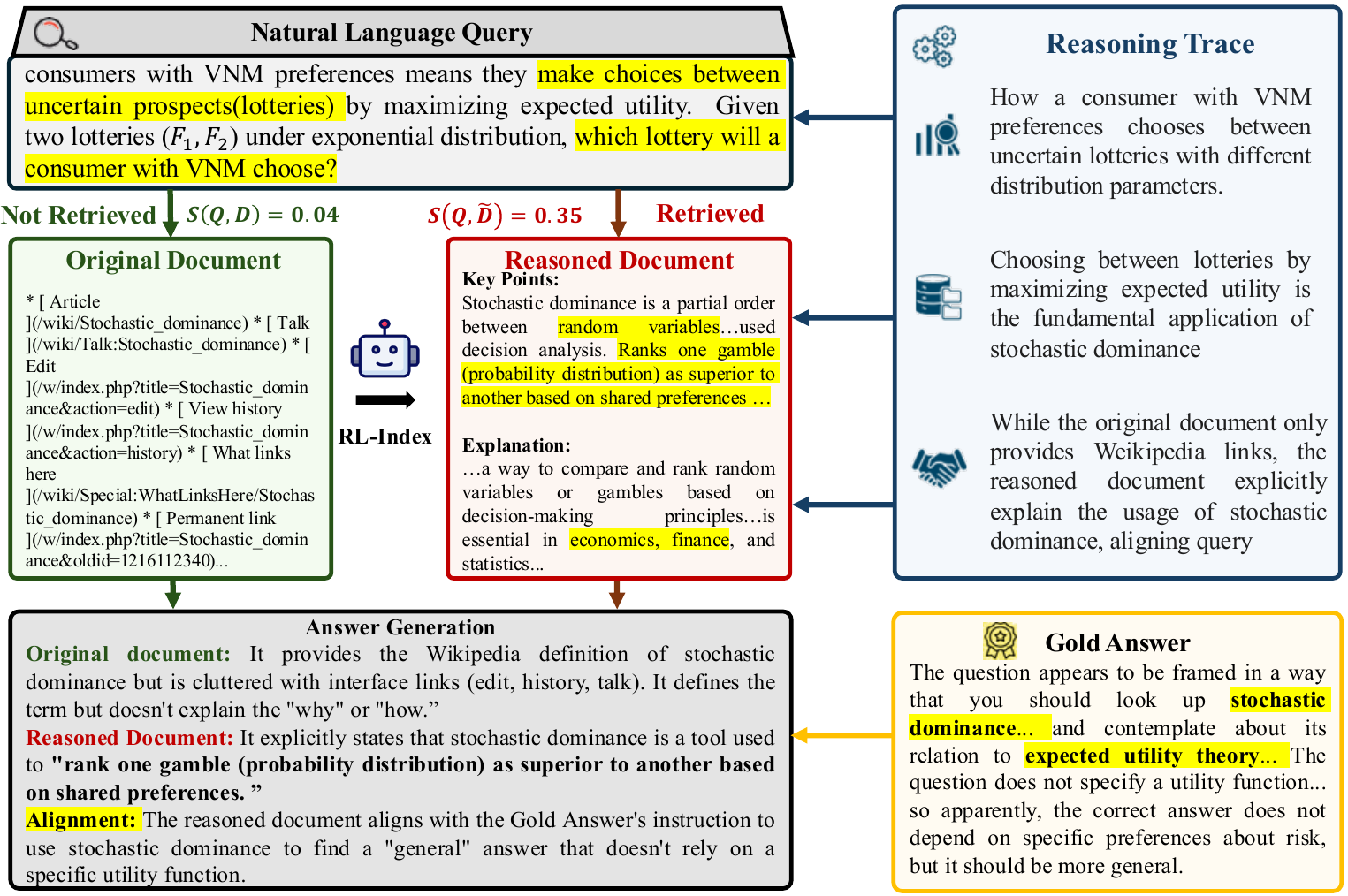}
    \caption{Retrieval and QA case study in the natural language domain.}
    \label{fig-retrieval-nl}
\end{figure*}

\begin{figure*}[t!]
    \centering
    \includegraphics[width=1\linewidth]{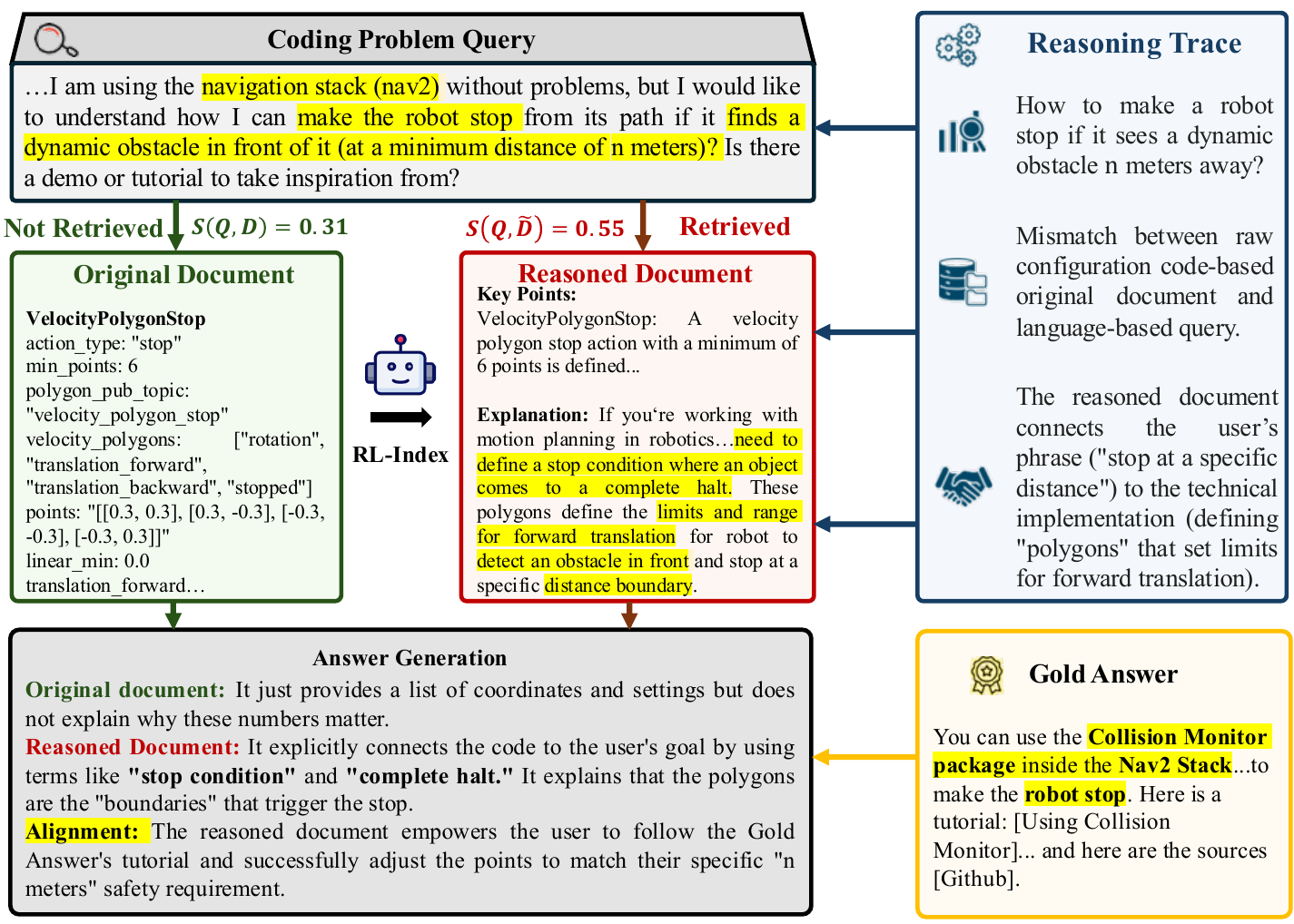}
    \caption{Retrieval and QA case study in the code domain.}
    \label{fig-retrieval-code}
\end{figure*}

To better understand why reasoned documents improve downstream QA performance, we additionally provide the gold answer and prompt an LLM to analyze how document reasoning enhances answer generation. 
As illustrated in Figure~\ref{fig-retrieval-nl}, although both the original and reasoned documents mention stochastic dominance, they differ significantly in how they support the query. The original document primarily focuses on the definition and includes external links without explaining how stochastic dominance can be applied, making it difficult to bridge the gap between the query and the gold answer. 
In contrast, the reasoned document explicitly explains that stochastic dominance can be used to rank gambles in the economics domain, directly aligning with the intent of the user’s query. This explicit connection to decision-making under uncertainty makes the reasoned document more closely aligned with the reasoning required to derive the gold answer.
Figure~\ref{fig-retrieval-code} presents another case study on a coding-related query. The user asks how to stop a robot at a minimum distance when detecting a dynamic obstacle using the Nav2 stack. Although the original document contains relevant configuration details (e.g., parameters for VelocityPolygonStop), it lists low-level coordinates and settings without explaining their purpose, making it difficult to connect these parameters to the user’s intent of \textit{stopping at a specific distance.}
In contrast, the reasoned document explains that the defined polygons act as boundaries that trigger the robot to stop or slow down, and clarifies how these boundaries relate to forward motion and obstacle detection. This reasoning aligns closely with the query’s intent and the gold answer, which references the Collision Monitor component in Nav2 for distance-aware stopping.
To summarize, by making the rationale more explicit, the reasoned document improves the usefulness of retrieved evidence for answer generation, leading to generating answers that are better aligned with the gold answer. 

\end{document}